\documentclass[aps,preprint,showpacs,nofootinbib]{revtex4-1}

\usepackage{graphicx}
\usepackage{slashed}
\usepackage{color}
\usepackage[colorlinks=true, linkcolor=blue, citecolor=blue, urlcolor=blue]{hyperref}

\begin{document}

\title{Heavy quarkonium production through the top quark rare decays via the channels involving flavor changing neutral currents}

\author{Juan-Juan Niu$^{1}$}
\email{niujj@cqu.edu.cn}
\author{Lei Guo$^{1}$}
\email{guoleicqu@cqu.edu.cn}
\author{Hong-Hao Ma$^{2}$}
\email{mahonghao.br@gmail.com}
\author{Shao-Ming Wang$^{1}$}
\email{smwang@cqu.edu.cn}

\address{$^{1}$ Department of Physics, Chongqing University, Chongqing 401331, P.R. China}
\address{$^{2}$ Faculdade de Engenharia de Guaratinguet\'a, Universidade Estadual Paulista, 12516-410, Guaratinguet\'a, SP, Brazil}

\date{\today}

\begin{abstract}

In the paper, we discuss the possibility of observation of heavy quarkoniums via the processes involving flavor changing neutral currents (FCNC). More explicitly, we systematically calculate the production of heavy charmonium and $(c\bar{b})$-quarkonium through the top quark semi-exclusive rare FCNC decays in the framework of the non-relativistic QCD (NRQCD) factorization theory. Our results show that the total decay widths $\Gamma_{t\to \eta_c} =1.20^{+1.04+1.14}_{-0.51-0.45}\times 10^{-16}$ GeV, $\Gamma_{t\to J/\psi} =1.37^{+1.03+1.30}_{-0.51-0.51}\times 10^{-16}$ GeV, $\Gamma_{t\to B_c}=2.06^{+0.17+0.91}_{-0.17-0.54}\times 10^{-18}$ GeV, and $\Gamma_{t\to B^*_c}=6.27^{+0.63+2.78}_{-0.62-1.64}\times 10^{-18}$ GeV, where the uncertainties are from variation of quark masses and renormalization scales. Even though the decay widths are small, it is important to make a systematic study on the production of charmonium and $(c\bar{b})$-quarkonium through the top-quark decays via FCNC in the Standard Model, which will provide useful guidance for future new physics research from the heavy quarkonium involved processes.  \\


\end{abstract}

\maketitle

\section{Introduction}

Since the discovery of the heavy quarkonium, the research on it attracts more and more attentions from theorists and experimentalists. As an important way to study the QCD mechanism, the production of heavy quarkonium is very useful for testing perturbative QCD (pQCD) theory~\cite{yellow1, yellow2, cms, cdf}. Many studies have been paid for them. For example, for the $B_c$ meson production, many studies have been done through not only the `direct' hadronic production~\cite{bc1, bc2, bc21, bc3}, but also its `indirect' production channels of top-quark~\cite{tbc1, tbc2}, $Z^0$-boson~\cite{zbc0, zbc1, zbc2, zbc3}, $W^{\pm}$-boson~\cite{wbc1, wbc2, w} and Higgs-boson~\cite{hpsai, hbc} decays in which sizable number of events can be detected at LHC or HL-LHC~\cite{cy1, cy2} which runs at the center-of mass energy $\sqrt{S}=14$ TeV with the current integrated luminosity of $3{~\rm ab}^{-1}$.

Being the heaviest fermion with a mass close to the electroweak symmetry breaking scale in standard model (SM), the top quark is helpful for analyzing the production of the heavy quarkonium and is also speculated to be a sensitive probe of new physics beyond the SM. A better understanding of those channels within the SM is helpful for judging whether there is really new physics, i.e. to deduct the SM background from the experimental data at a high confidence level such that to determine the right ranges for the new physics parameters. Following the top quark dominant decay channel, $t\rightarrow b W^+$, it has been pointed out that sizable $B^{-}_c$ mesons can be produced via the channel, $t\to |(b\bar{c})[n]\rangle +c +W^{+}$~\cite{tbc1, tbc2}, where $[n]$ stands for the $(b\bar{c})$-quarkonium state via the velocity scaling rule of the non-relativistic QCD (NRQCD) theory~\cite{nrqcd}.

The heavy quarkonium ($B^-_{c}$, $\eta_c$ and etc.) may also be produced via the top-quark decays through the flavor changing neutral current (FCNC) processes, i.e. $t\to |(c\bar{Q})[n]\rangle + Q+ Z^{0}$ with $Q$ equals to $c$ or $b$ respectively. The FCNC processes involving heavy hadrons are of significant interests and allow stringent tests of our current understanding of particle physics. The Glashow-Iliopoulos-Maiani (GIM) mechanism~\cite{gim} forbids its production at the tree level and covers important information in the loop structure. There are many studies focused on the top-quark rare decays via FCNC in the SM~\cite{tcz, tj, jnl} and other new models like the two-Higgs-doublet models (2HDM)~\cite{tcz}, the minimal supersymmetric model (MSSM)~\cite{jcj}, the Topcolor-assisted Technicolor Model (TC2)~\cite{xcg} and other models~\cite{pr}. These researchs confirmed that FCNC processes could be unambiguous small but also could provide a useful window in the quest for new physics signals. Thus to make a systematic study on the production of charmonium and $c\bar{b}$-quarkonium through the top-quark decays via the FCNC in the SM is requisite, it will provide useful guidance for future new physics research from the heavy quarkonium involved processes. As will be shown later, the decay width via FCNC is generally small and the contribution from the $P$-wave states is relatively smaller than that of the $S$-wave states. In the present paper, we shall only make a detailed discussion on the production of two color-singlet $S$-wave states $^1S_0$ and $^3S_1$.

The remaining parts of the paper are organized as follows. In Sec.II, we present the calculation technology for the production of heavy quarkonium through the top-quark rare decays via FCNC. Numerical results for total and differential decay widths, together with their uncertainties, are presented in Sec.III. Sec.IV is reserved for a summary.

\section{Calculation Technology}

\begin{figure}[htb]
\includegraphics[width=0.25\textwidth]{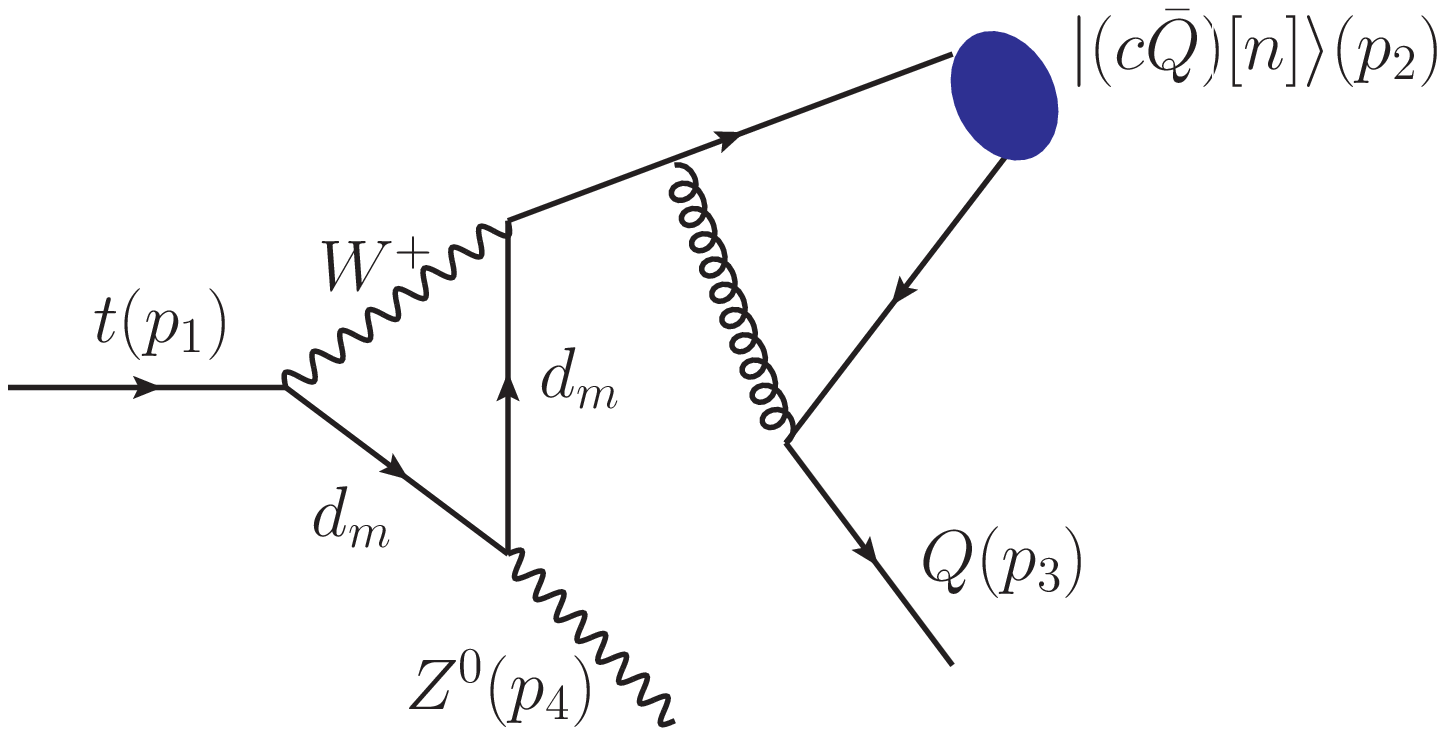}
\includegraphics[width=0.25\textwidth]{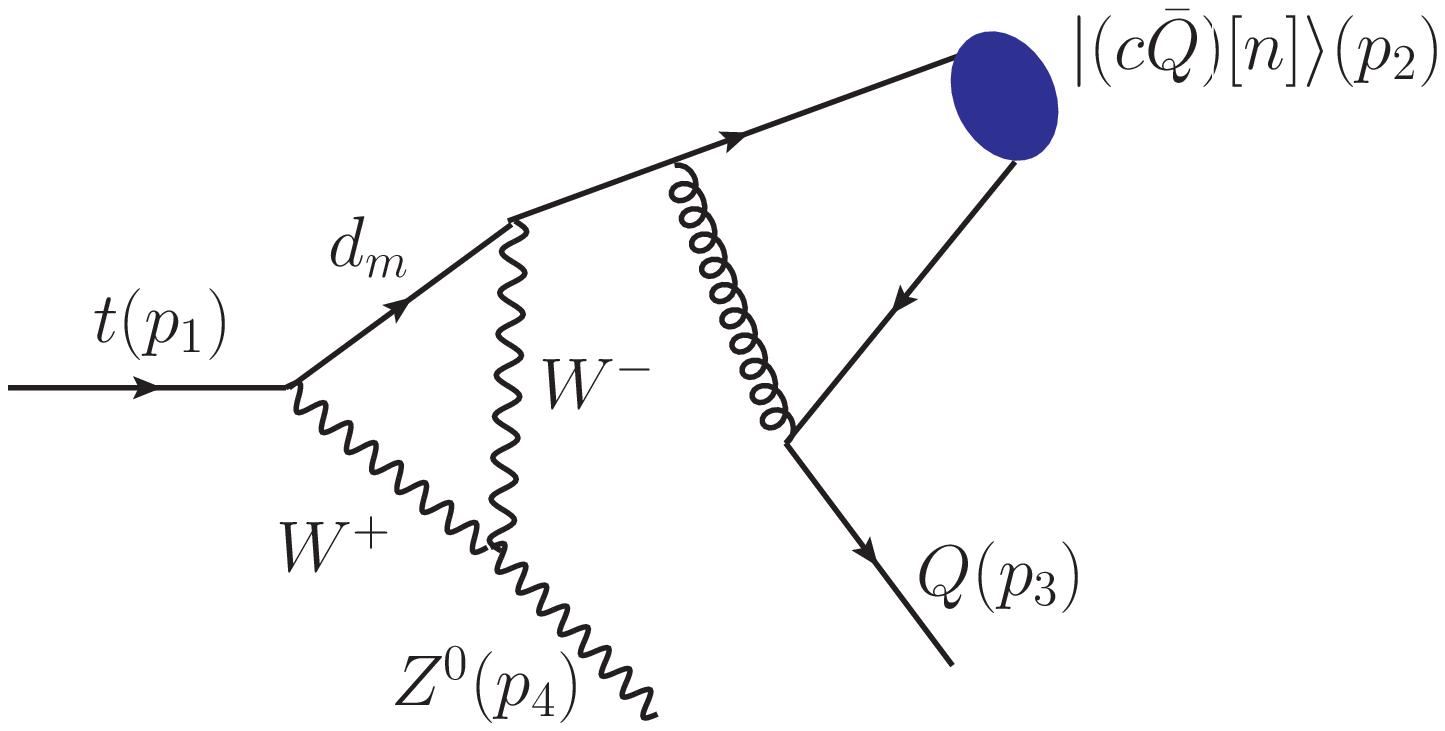}
\includegraphics[width=0.25\textwidth]{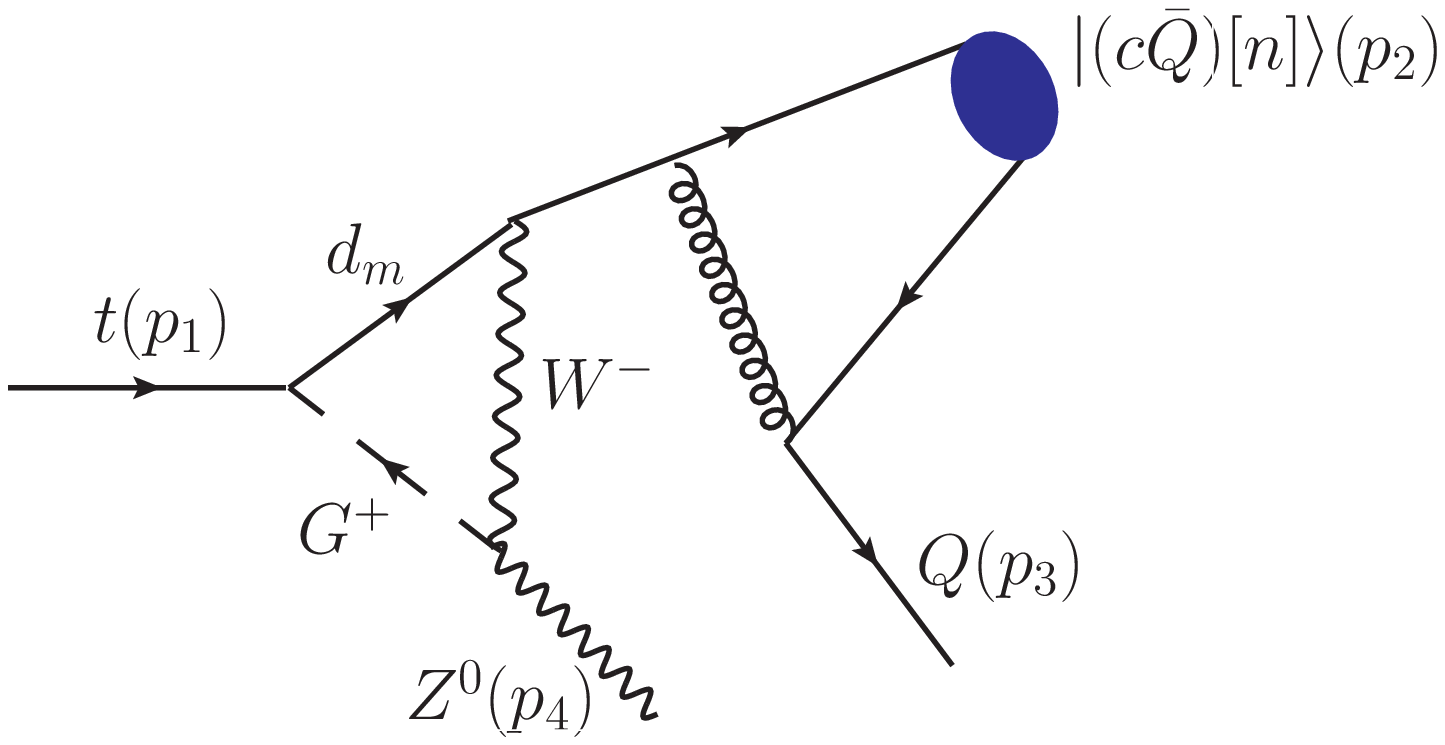}
\includegraphics[width=0.25\textwidth]{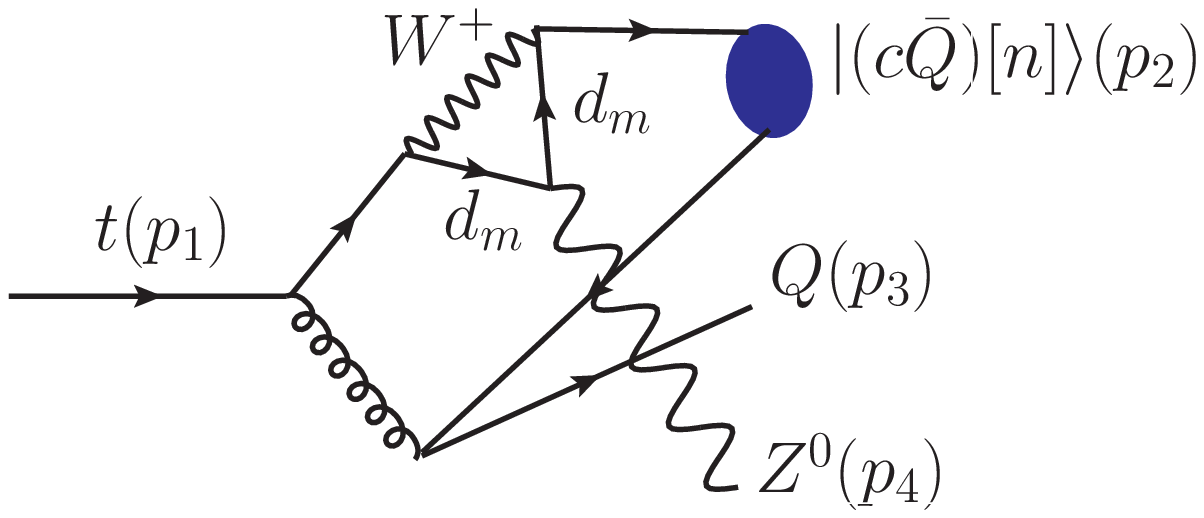}
\includegraphics[width=0.25\textwidth]{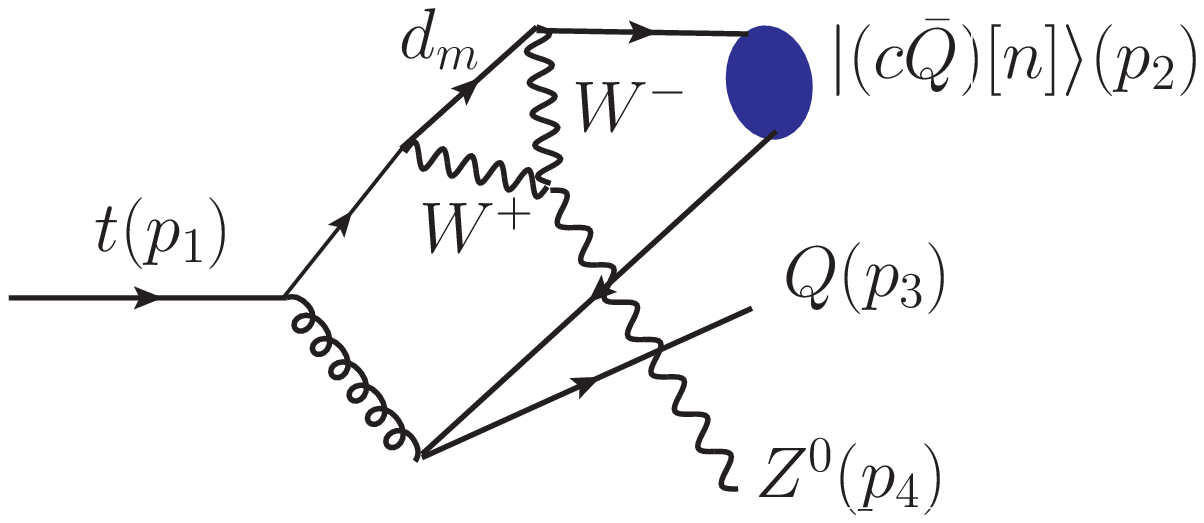}
\includegraphics[width=0.25\textwidth]{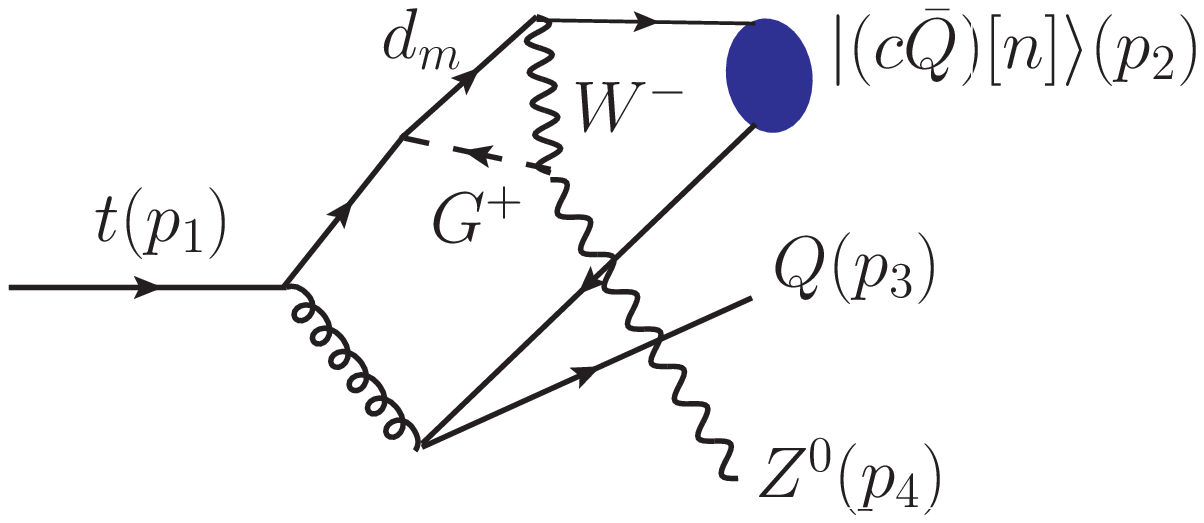}
\includegraphics[width=0.24\textwidth]{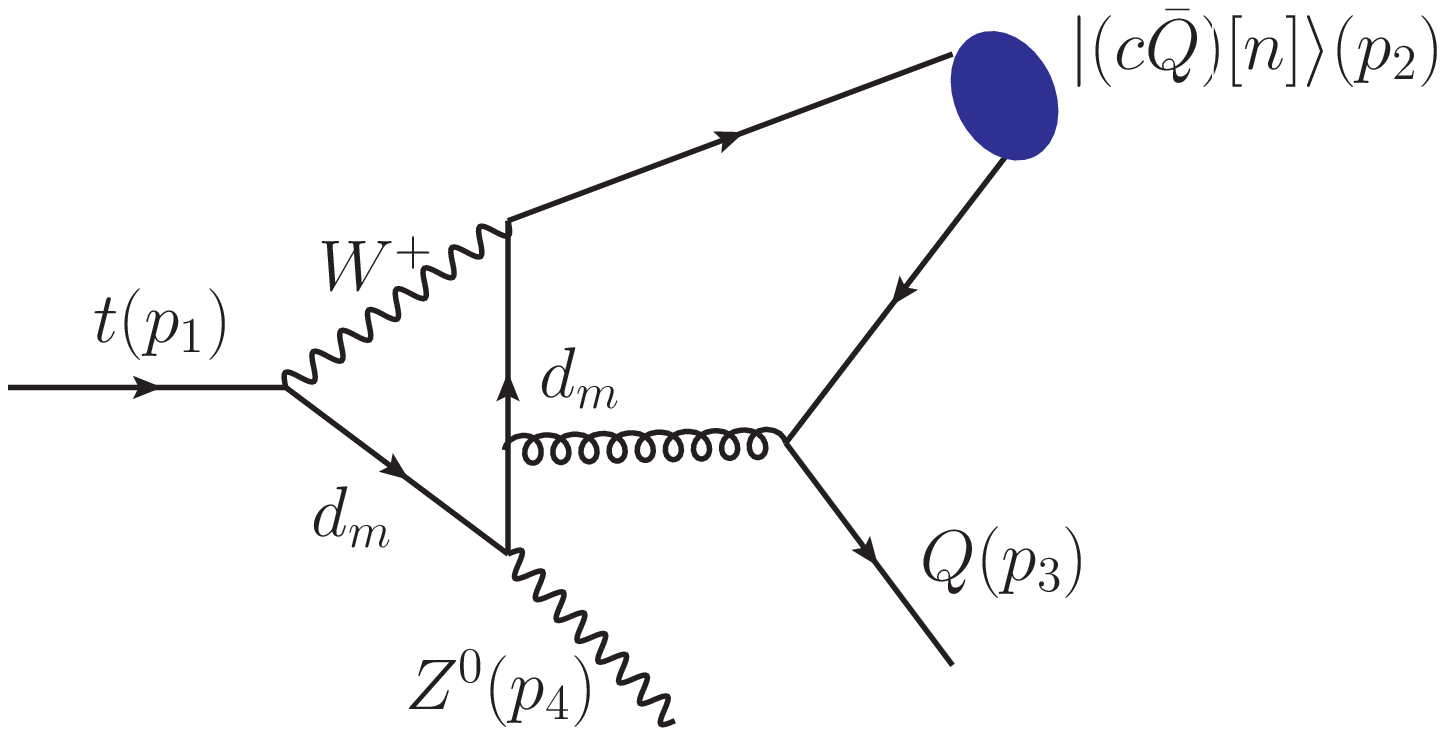}
\includegraphics[width=0.24\textwidth]{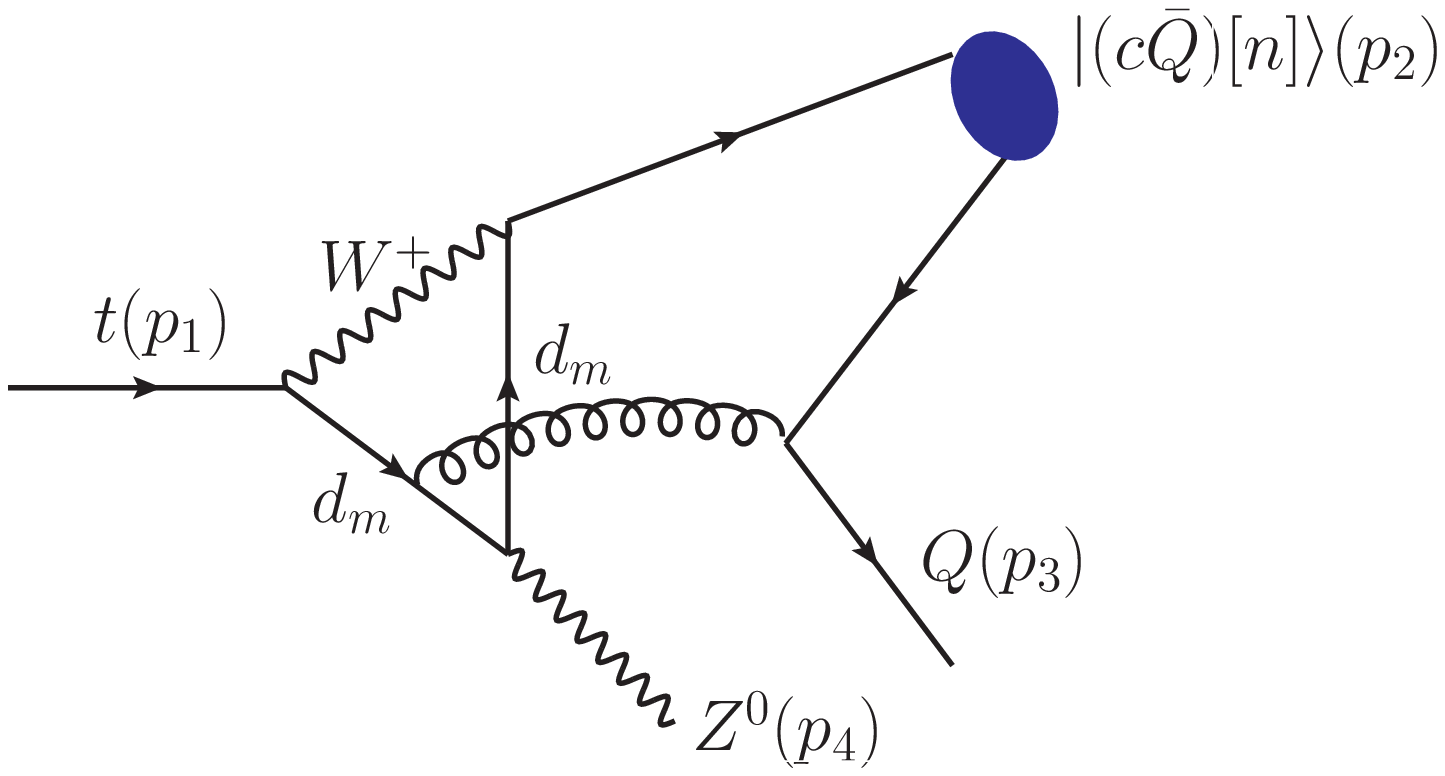}
\includegraphics[width=0.24\textwidth]{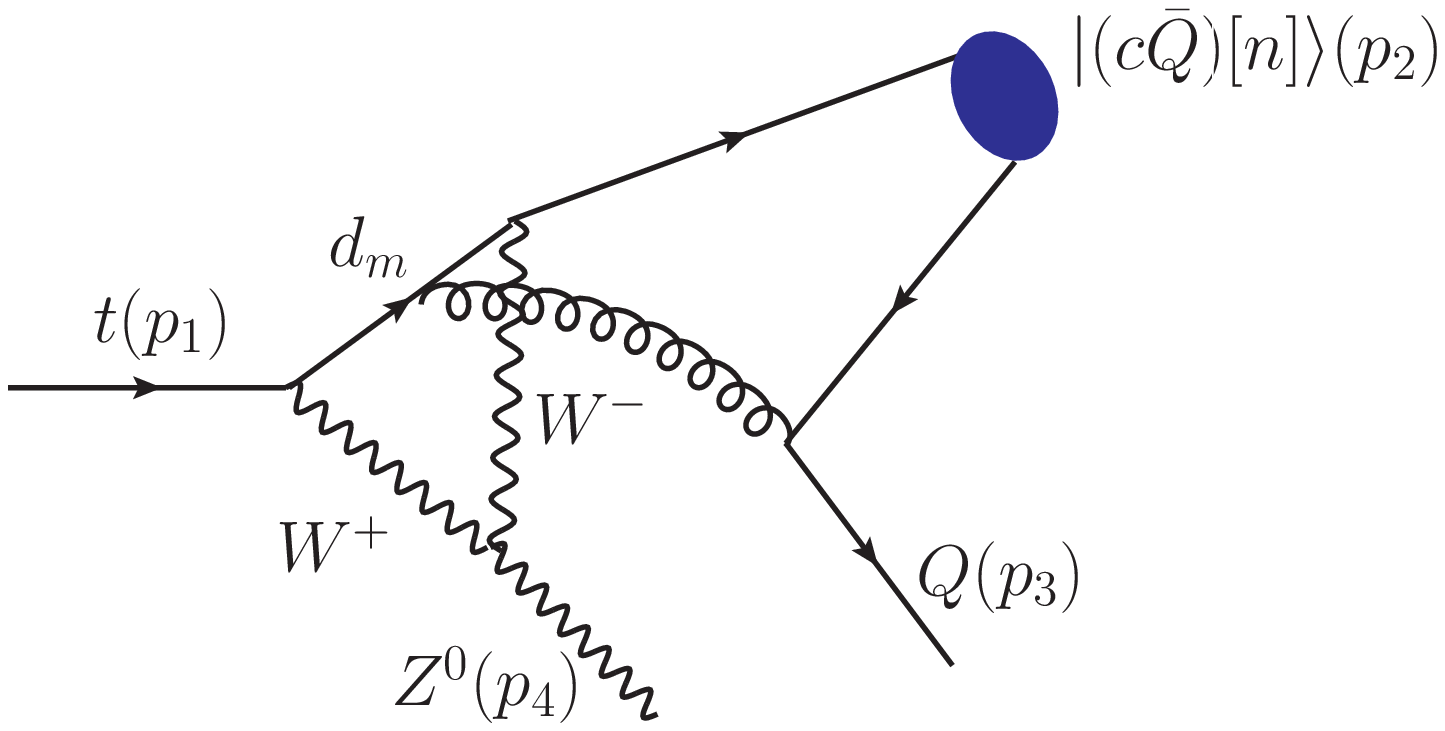}
\includegraphics[width=0.24\textwidth]{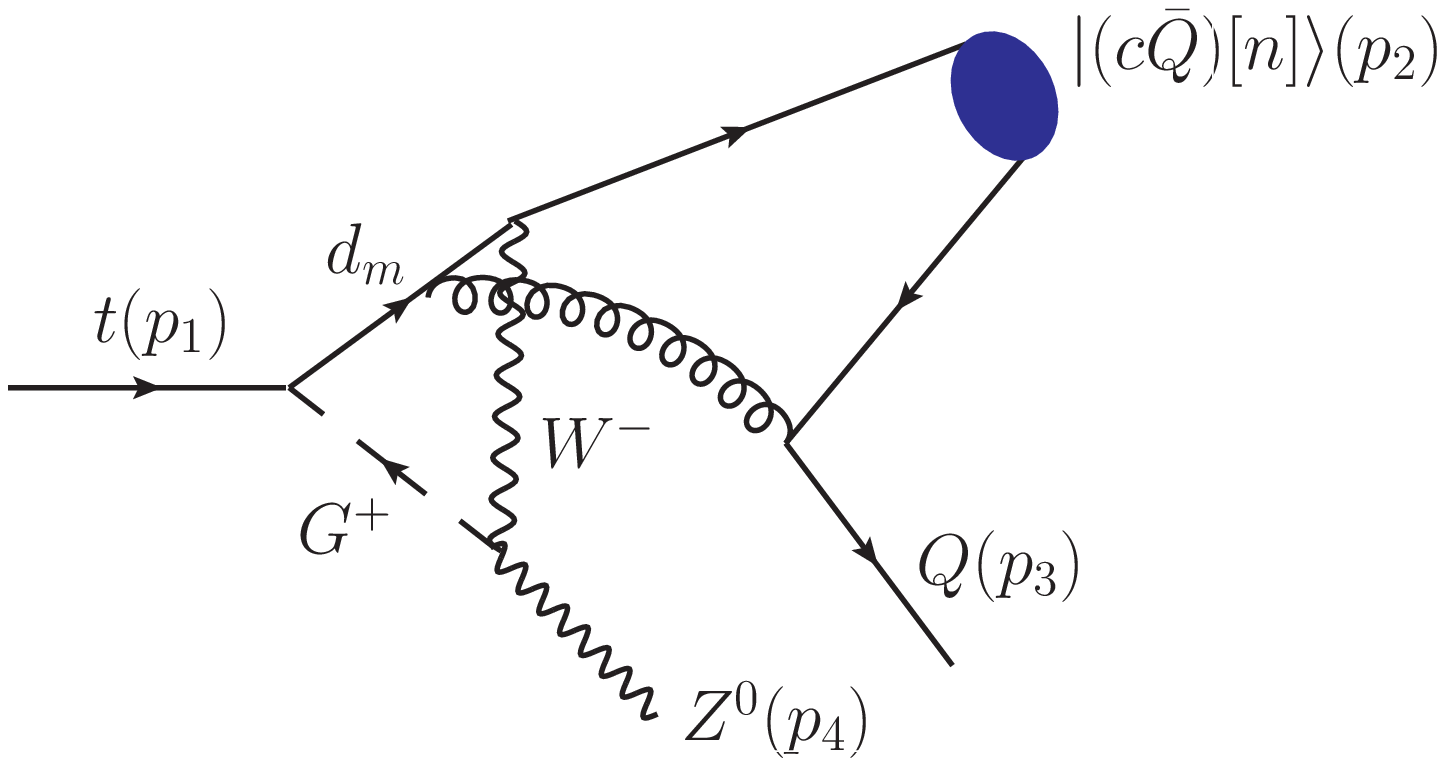}
\caption{Feynman diagrams for the FCNC production channel, $t(p_1)\to |(c\bar{Q})[n]\rangle(p_2) + Q(p_3) + Z^{0}(p_4)$, where $n$ stands for a series of Fock states. $d_m$ denotes as the generation of down-type quark with mass $m_{d_m}$.} \label{feyn1}
\end{figure}

The FCNC kernel of the top-quark decay is $t\to c Z^0$, and the charmonium and the $(c\bar{b})$-quarkonium production via FCNC is through the process
\begin{equation}
t(p_1) \to |(c\bar{Q})[n]\rangle(p_2) + Q(p_3) + Z^0(p_4),
\end{equation}
where $Q$ stands for $c$ or $b$, $p_i~(i = 1,2,3,4)$ represent the four-momenta of initial and final state particles, respectively. Feynman diagrams for the production of heavy quarkonium via FCNC are depicted in Fig.(\ref{feyn1}), where the $t\to cZ^{0}$ is realized via a weak interaction loop. Compared to $b$ and $s$ quarks, the $d$ quark can be ignored for its small mass and the $\mathrm{CKM}(1,3)$ is only $0.009$. Because the intermediate gluon should be hard enough to generate a $c\bar{c}$ pair or a $b\bar{b}$ pair, those processes are pQCD calculable. The specific momenta of the two constitute quarks in $(c\bar{Q})$-quarkonium are $p_{21}$ and $p_{22}$:
\begin{equation}
p_{21} = \frac{m_c}{M}{p_2} + q ,\;\;\;p_{22} = \frac{m_{Q}}{M}{p_2} - q,
\end{equation}
where $q$ stands for the relative momentum between the two constituent quarks. The Quarkonium mass $M\simeq m_c + m_{Q}$ is adopted to ensure the gauge invariance of the hard scattering amplitude.

The decay width of the process $t \to |(c\bar{Q})[n]\rangle + Q + Z^{0}$ can be written in the following factorized form
\begin{equation}
\Gamma=\sum_{n} \hat\Gamma(t\to |(c\bar{Q})[n]\rangle + Q+Z^{0}) \langle{\cal O}^H[n] \rangle ,
\end{equation}
where $n$ means a series of Fock states. Contributions from the color-octet states or the $P$-wave states are generally smaller than that from the color-singlet $S$-wave states, which are about $10\%$ of the ground states via a general velocity scaling rule~\cite{nrqcd}. Thus in the present paper, we shall consider the color-singlet $S$-wave states' contributions. The non-perturbative matrix element $\langle{\cal O}^{H}[n]\rangle$ describes the hadronization process of a perturbative $(c\bar{Q})$ pair into an observable hadronic state. The color-singlet ones can be computed through potential models~\cite{pot1, pot2, pot3, pot4, pot5, pot6}, e.g. the color-singlet $S$-wave states are related to the wavefunction at the origin, $\Psi_S(0)^2=R_S(0)^2/{4\pi}$. The decay width $\hat\Gamma$ represents the short-distance coefficients which can be calculated perturbatively
\begin{equation}
\hat\Gamma =\int \frac{1}{2m_t} \overline{\sum} |M|^{2} d\Phi_3,
\end{equation}
where the symbol $\overline{\sum}$ means to sum over the color and spin of final-state particles and to average over the spin and color of initial-state top quark. $d\Phi_3$ is the three-body phase space which can be written as
\begin{equation}
d{\Phi_3}=(2\pi)^4 \delta^{4}\left(p_1 - \sum_{f=2}^4 p_{f}\right)\prod_{f=2}^4 \frac{d^3{\vec{p}_f}}{(2\pi)^3 2p_f^0},
\end{equation}
It is helpful to get the differential distributions, $d\Gamma/ds_{ij}$ and $d\Gamma/d\cos\theta_{ij}$, for experimental studies, where the invariant masses $s_{ij}=(p_i+p_j)^2$ and $\theta_{ij}$ is the angle between $\vec{p}_i$ and $\vec{p}_j$ for $i,j=2,3,4$.

The amplitude can be generally expressed as
\begin{equation}
i M_{ss'}[n] = {\cal{C}}\; {\bar {u}_{s i}}({p_3}) \sum\limits_{l = 1}^{m} {{\cal A}_l[n] } {u_{s' j}}({p_1}),
\end{equation}
where $m=10$ stands for the number of Feynman diagrams of this processes, $s$ and $s'$ are spin indices, $i$ and $j$ are color indices of the outgoing $Q$ quark and the initial top quark, respectively. The color factor ${\cal C}$ for the color-singlet production is $\frac{4}{3 \sqrt{3} }\delta_{ij}$. The amplitude ${\cal A}_{l}[n]$ for each hadronic state can be read out from Feynman diagrams in Fig.(\ref{feyn1}). It is worth mentioning that $v(p_{22})\bar{u}(p_{21})$ for $(c\bar{Q})$-quarkonium in ${\cal A}_l[n]$ must be replaced by the projector $\Pi_{p_2}[n]$ for each corresponding Fock state. And the projector $\Pi_{p_2}[n]$ for the spin-singlet or spin-triplet $S$-wave states can be written as~\cite{projector}:
\begin{eqnarray}
\Pi_{p_2}[n] &=& \frac{1}{2\sqrt{M}}\epsilon[n](\slashed{p}_{2}+ M).
\end{eqnarray}
where $\epsilon[^1S_0]=\gamma_5$ and $\epsilon[^3S_1]=\slashed{\epsilon}$ with $\epsilon^\rho$ is the polarization vector of $^3S_1$ state.

As for the present considered one-loop triangle integrals with three internal masses, it is noted that there is no ultra-violet divergence~\cite{oneloop}, thus we can get the finite results by directly performing the loop integrals. More explicitly, the amplitudes $A_{l}$ are given in Appendix A.

\section{Numerical Results}

We use FeynArts 3.9~\cite{feynarts} to generate amplitudes and the modified FormCalc 7.3/LoopTools 2.1~\cite{formcalc} to do the algebraic and numerical calculations. We set the typical renormalization scale $\mu_R$ to be $2m_c$ ($2m_b$) for the production of charmonium ($(c\bar{b})$-quarkonium) accordingly, leading to $\alpha_s(2m_c)=0.259$ and $\alpha_s(2m_b)=0.181$. Because the wavefunction at the zero is an overall factor and its uncertainty can be conventionally discussed when we know its exact values, thus we shall directly take the wavefunction at the zero to be the one derived from the QCD (Buchmuller-Type) potential model~\cite{pot6}. We set the masses of the ground states charmonium and ($c\bar{b}$)-quarkonium as 3~GeV~\cite{Aaij:2014bga,Anashin:2014wva} and 6.4~GeV~\cite{Abe:1998wi,Aaij:2016qlz,Aaij:2014asa} by default. As a summary, the relevant input parameters are as follows:
\begin{eqnarray}
&m_{Z}=91.1876~{\rm GeV}, ~~m_{W}=80.385~{\rm GeV}, ~~m_t = 173.0~{\rm GeV},&\nonumber\\
&m_c=1.50~{\rm GeV}, ~~m_b=4.90~{\rm GeV}, ~~m_s=0.101~{\rm GeV},&\nonumber\\
&|R_S(c\bar{c})(0)|^2=0.810\;{\rm GeV}^3, ~~|R_S(c\bar{b})(0)|^2=1.642\;{\rm GeV}^3, ~~G_F=1.1663787\times10^5.& \nonumber
\end{eqnarray}

\subsection{The charmonium and  $(c\bar{b})$-quarkonium production via FCNC}

Total decay width for the process $t\to cZ^{0}$  is $9.59\times 10^{-13}$ GeV£¬ which is small due to the strong GIM suppression from the small values of the internal quark masses $m_{b, s, d}$. As a subtle point, contribution from the ${d}$ quark loop is negligible due to small CKM matrix element $|V_{td}|$ and its small mass.

\begin{table}[htb]
\begin{tabular}{|c|c|c|c|}
\hline
~~$t\rightarrow |(c\bar{Q})[n]\rangle$~~ & ~~$\Gamma$ (GeV)~~ & ~~$R$~~ \\
\hline\hline
$t\rightarrow\eta_c$ & $1.20\times 10^{-16}$ & $1.25\times10^{-4}$ \\
\hline
$t\rightarrow J/\psi$ & $1.37\times 10^{-16}$ & $1.43\times10^{-4}$ \\
\hline\hline
$t\rightarrow B_c $ &  $2.06\times 10^{-18}$ & $2.15\times 10^{-6}$ \\
$t\rightarrow B^*_c$ & $6.27\times 10^{-18}$ & $6.54\times 10^{-6}$ \\
\hline
\end{tabular}
\caption{The decay widths and the corresponding branching ratios for the $(c\bar{Q})$-quarkonium production via the channel $t\to |(c\bar{Q})[n]\rangle +Q+ Z^{0}$. The ratio $R={\Gamma_{t\rightarrow |(c\bar{Q})[n]\rangle}}/{\Gamma_{t\rightarrow cZ^{0}}}$. }
\label{tabcc}
\end{table}

The decay width and corresponding branching ratios for the production of the $(c\bar{Q})$-quarkonium through the channel $t \to |(c\bar{Q})[n]\rangle +Q +Z^{0}$ via FCNC are listed in Table~\ref{tabcc}. Table~\ref{tabcc} shows the decay width of the charmonium production is almost two orders of magnitude  larger than that of the $(c\bar{b})$-quarkonium production.

\begin{figure}[htb]
\includegraphics[scale=0.6]{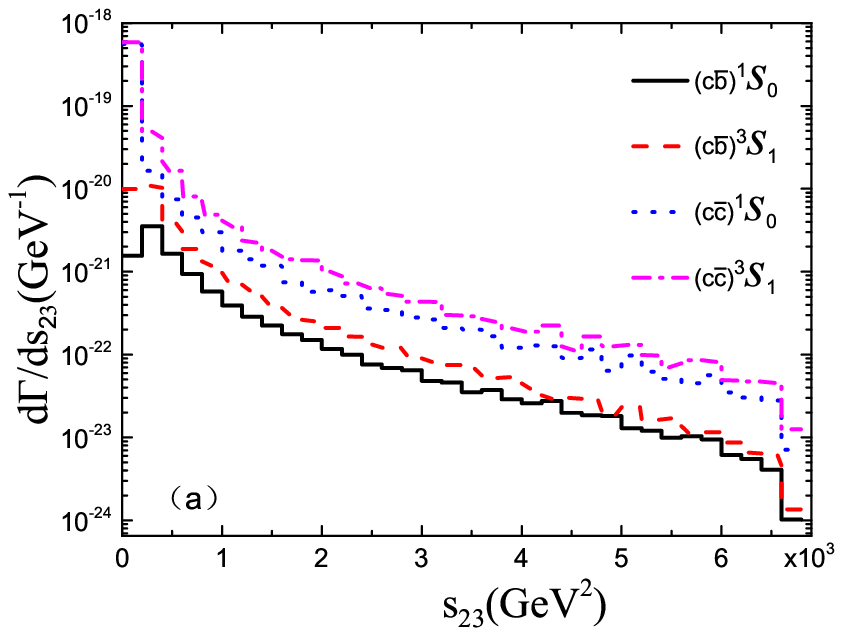}
\includegraphics[scale=0.6]{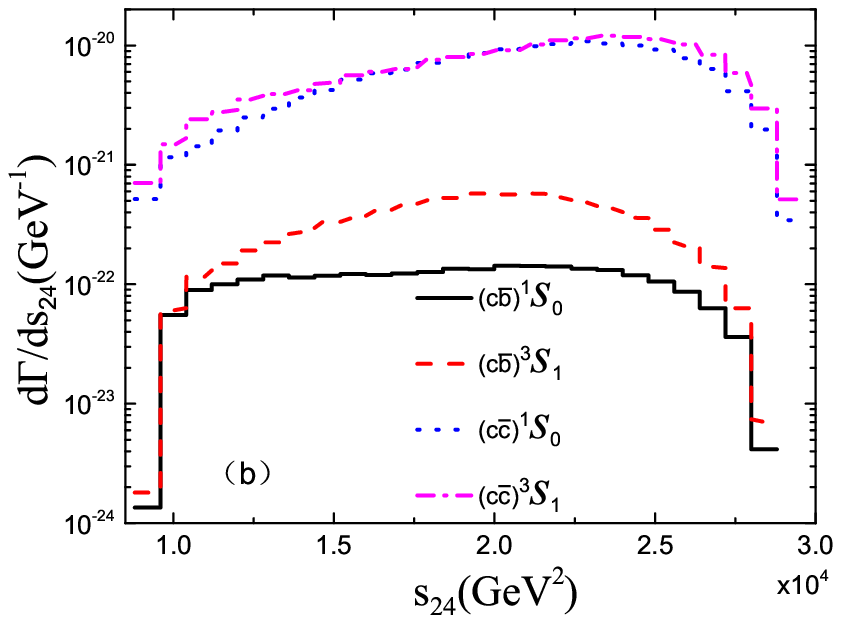}
\includegraphics[scale=0.6]{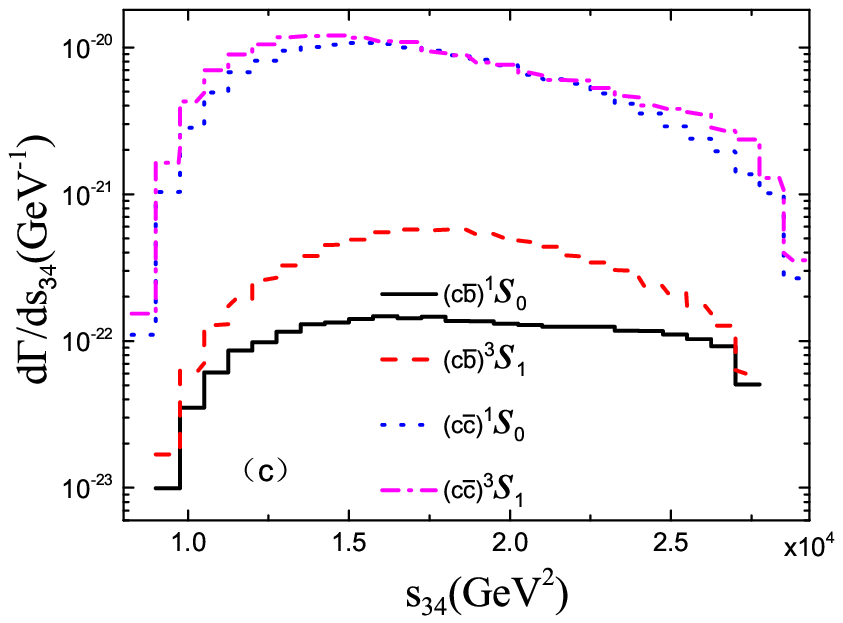}
\caption{The differential decay widths $d\Gamma/ds_{23}$ (a), $d\Gamma/ds_{24}$ (b), and $d\Gamma/ds_{34}$ (c) for $ t\rightarrow |(c\bar{Q})[n]\rangle +Q+Z^{0}$. } \label{ds}
\end{figure}

\begin{figure}[htb]
\includegraphics[scale=0.6]{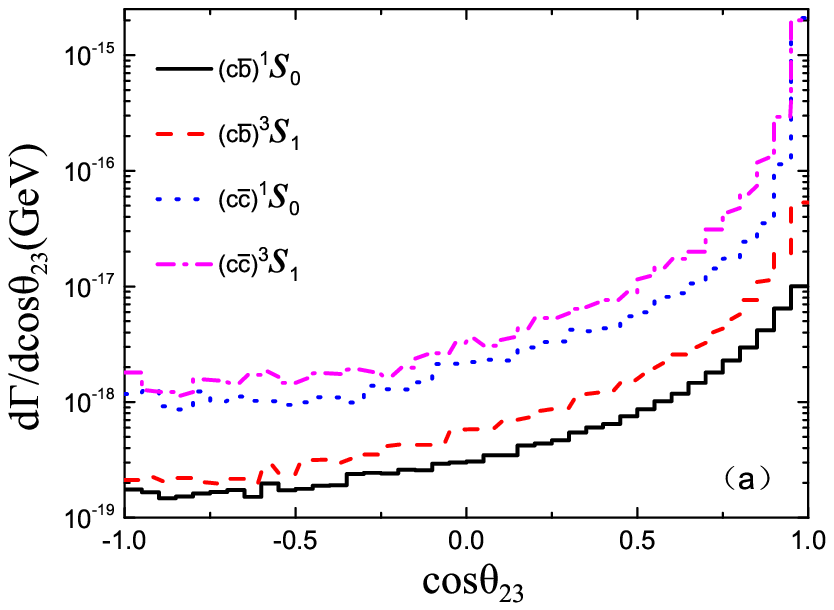}
\includegraphics[scale=0.6]{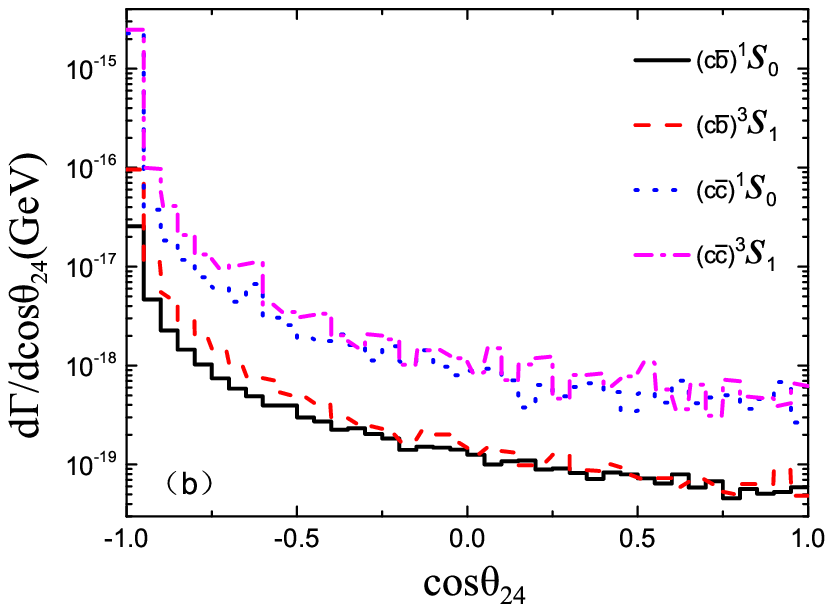}
\includegraphics[scale=0.6]{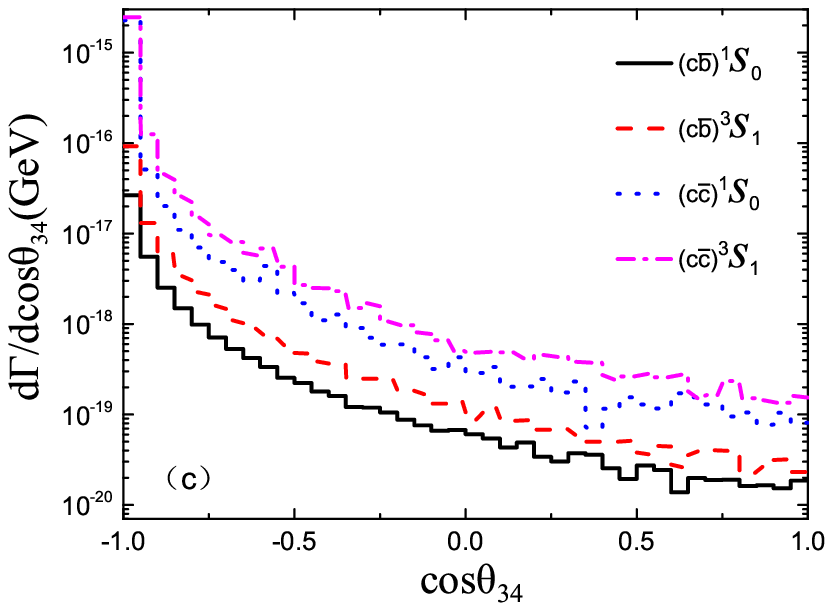}
\caption{The differential decay widths $d\Gamma/d\cos\theta_{23}$ (a), $d\Gamma/d\cos\theta_{24}$ (b) and $d\Gamma/d\cos\theta_{34}$ (c) for $ t\rightarrow |(c\bar{Q})[n]\rangle +Q+Z^{0}$. }
\label{dcos}
\end{figure}

We present the differential distributions over the invariant masses $d\Gamma/ds_{23}$, $d\Gamma/ds_{24}$ and $d\Gamma/ds_{34}$ and the differential distributions over the angles $d\Gamma/d\cos\theta_{23}$, $d\Gamma/d\cos\theta_{24}$ and $d\Gamma/d\cos\theta_{34}$ between the final particles for the $|(c\bar{Q})[n]\rangle$ production in Figs.(\ref{ds}, \ref{dcos}), respectively. In Figs.(\ref{ds}) the sharp peaks in low region of $d\Gamma/ds_{23}$ indicate the largest contribution emerges when the heavy quarkonium moves along with the same direction of the outgoing quark but with the opposite direction of the outgoing $Z^0$ boson. This feature is consistent with angle distributions in Figs.(\ref{dcos}).

\subsection{Uncertainties for the charmonium and  $(c\bar{b})$-quarkonium production via FCNC}

There are uncertainties from different choices of quark masses, renormalization scale and wavefunction uncertainties. In this subsection, we discuss the uncertainties from the quark masses and the renormalization scale.

\begin{table}[htb]
\begin{tabular}{|c||c|c|c|}
\hline
~~~~~        & ~~$m_c=1.25$ GeV~~   & ~~$m_c=1.50$ GeV~~   & ~~$m_c=1.75$ GeV~~  \\
\hline\hline
$\Gamma_{|(c\bar{c})[^1S_0]\rangle}$ & $2.24\times 10^{-16}$  & $1.20\times 10^{-16}$  & $0.69\times 10^{-16}$  \\
\hline
$\Gamma_{|(c\bar{c})[^3S_1]\rangle}$ &  $2.40\times 10^{-16}$  & $1.37\times 10^{-16}$   & $0.86\times 10^{-16}$   \\
\hline\hline
$\Gamma_{|(c\bar{b})[^1S_0]\rangle}$ &  $2.06\times 10^{-18}$  & $2.06\times 10^{-18}$  & $2.06\times 10^{-18}$  \\
\hline
$\Gamma_{|(c\bar{b})[^3S_1]\rangle}$ & $6.53\times 10^{-18}$  & $6.27\times 10^{-18}$  &  $6.06\times 10^{-18}$  \\
\hline
\end{tabular}
\caption{Uncertainties of the decay width for the process $t\rightarrow |(c\bar{Q})[n]\rangle +Q +Z^0$ by varying $m_c\in[1.25,1.75]$ GeV.}
\label{tabuncernmc}
\end{table}

\begin{table}[htb]
\begin{tabular}{|c||c|c|c|}
\hline ~~~~    & ~~$m_b=4.50$ GeV~~   & ~~$m_b=4.90$ GeV~~   & ~~$m_b=5.30$ GeV~~  \\
\hline \hline
$\Gamma_{|(c\bar{c})[^1S_0]\rangle}$ &  $0.82\times 10^{-16}$   & $1.20\times 10^{-16}$    & $1.70\times 10^{-16}$   \\
\hline
$\Gamma_{|(c\bar{c})[^3S_1]\rangle}$ & $0.98\times 10^{-16}$  & $1.37\times 10^{-16}$   &  $ 1.88\times 10^{-16}$  \\
\hline
\hline
$\Gamma_{|(c\bar{b})[^1S_0]\rangle}$ &  $1.89\times 10^{-18}$   & $2.06\times 10^{-18}$    & $ 2.23\times 10^{-18}$   \\
\hline
$\Gamma_{|(c\bar{b})[^3S_1\rangle}$ & $5.65\times 10^{-18}$  & $6.27\times 10^{-18}$   &  $ 6.90\times 10^{-18}$  \\
\hline
\end{tabular}
\caption{Uncertainties of the decay width for the process $t\rightarrow |(c\bar{Q})[n]\rangle +Q +Z^0$ by varying $m_b\in[4.50,5.30]$ GeV.}
\label{tabuncernmb}
\end{table}

\begin{table}[htb]
\begin{tabular}{|c||c|c|c|}
\hline
~~~~        & ~~$m_t=169.0$ GeV~~   & ~~$m_t=173.0$ GeV~~   & ~~$m_t=177.0$ Gev~~  \\
\hline\hline
$\Gamma_{|(c\bar{c})[^1S_0]\rangle}$ & $ 1.15\times 10^{-16}$ & $1.20\times 10^{-16}$   & $ 1.25\times 10^{-16}$ \\
\hline
$\Gamma_{|(c\bar{c})[^3S_1]\rangle}$ & $ 1.32\times 10^{-16}$  & $1.37\times 10^{-16}$  & $ 1.45\times 10^{-16}$ \\
\hline\hline
$\Gamma_{|(c\bar{b})[^1S_0]\rangle}$ & $2.05\times 10^{-18}$ & $2.06\times 10^{-18}$  & $2.08\times 10^{-18}$ \\
\hline
$\Gamma_{|(c\bar{b})[^3S_1]\rangle}$ & $5.71\times 10^{-18}$ & $6.27\times 10^{-18}$ & $6.88\times 10^{-18}$   \\
\hline
\end{tabular}
\caption{Uncertainties of the decay width for the process $t\rightarrow |(c\bar{Q})[n]\rangle +Q +Z^0$ by varying $m_t\in[169.0,177.0]$ GeV.}
\label{tabuncernmt}
\end{table}

In Tables~\ref{tabuncernmc}, \ref{tabuncernmb} and \ref{tabuncernmt}, we present the uncertainties caused by $m_c$, $m_b$ and $m_t$ within the range of $m_c=1.50\pm0.25$ GeV, $m_b=4.90\pm0.40$ GeV and $m_t=173.0\pm4.0$ GeV. When varying one mass parameter, the other two parameters are fixed to be their central values. Tables~\ref{tabuncernmc}, \ref{tabuncernmb} and \ref{tabuncernmt} indicate that the mass uncertainties are large. The decay width for the production of both charmonium and $(c\bar{b})$-quarkonium will increase $1\%\sim10\%$ with  the increment of $m_t$. For the charmonium production, its decay width decreases with the increment of $m_c$ and increases with the increment of $m_b$. For the production of $(c\bar{b})$-quarkonium, the decay width increases slower with the increment of $m_b$. The total decay widthes with mass uncertainties are
\begin{eqnarray}
\Gamma_{t\to \eta_c} &=& 1.20^{+1.04}_{-0.51}\times 10^{-16}\; {\rm GeV}, \\
\Gamma_{t\to J/\psi}  &=& 1.37^{+1.03}_{-0.51}\times 10^{-16}\; {\rm GeV},  \\
\Gamma_{t\to B_c}   &=& 2.06^{+0.17}_{-0.17}\times 10^{-18}\;  {\rm GeV}, \\
\Gamma_{t\to B^*_c}&=& 6.27^{+0.63}_{-0.62}\times 10^{-18}\;  {\rm GeV},
\end{eqnarray}
where the uncertainties from various quark masses are summed up in quadrature.

\begin{table}[htb]
\begin{tabular}{|c||c|c|c|}
\hline ~~~~    & ~$\mu_R$~   & ~~$\frac{1}{2}\mu_R$~~   & ~~2$\mu_R$~~  \\
\hline \hline
$\Gamma_{|(c\bar{c})[^1S_0]\rangle}$ & $1.20\times10^{-16}$   & $2.34\times10^{-16}$   & $0.75\times10^{-16}$  \\
\hline
$\Gamma_{|(c\bar{c})[^3S_1]\rangle}$ &$1.37\times10^{-16}$  & $2.67\times10^{-16}$  &  $0.86\times10^{-16}$  \\
\hline\hline
$\Gamma_{|(c\bar{b})[^1S_0]\rangle}$ & $2.06\times10^{-18}$   & $2.97\times10^{-18}$   & $1.52\times10^{-18}$  \\
\hline
$\Gamma_{|(c\bar{b})[^3S_1]\rangle}$ & $6.27\times10^{-18}$  & $9.05\times10^{-18}$  &  $4.63\times10^{-18}$  \\
\hline
\end{tabular}
\caption{Scale uncertainties of the decay width for the process $t\to |(c\bar{Q})[n]\rangle +Q +Z^0$ by varying the typical renormalization scale $\mu_R$ from $\frac{1}{2}\mu_R$ to $2\mu_R$. $\mu_R=2m_c$ for charmonium and $\mu_R=2m_b$ for $(c\bar{b})$-quarkonium. }
\label{tabmu}
\end{table}

\begin{figure}[htb]
\includegraphics[scale=0.6]{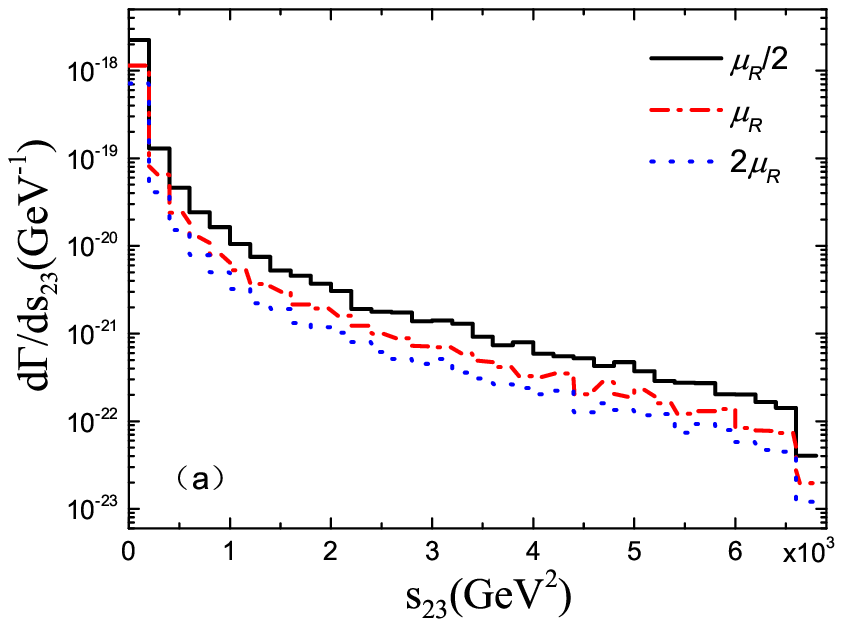}
\includegraphics[scale=0.6]{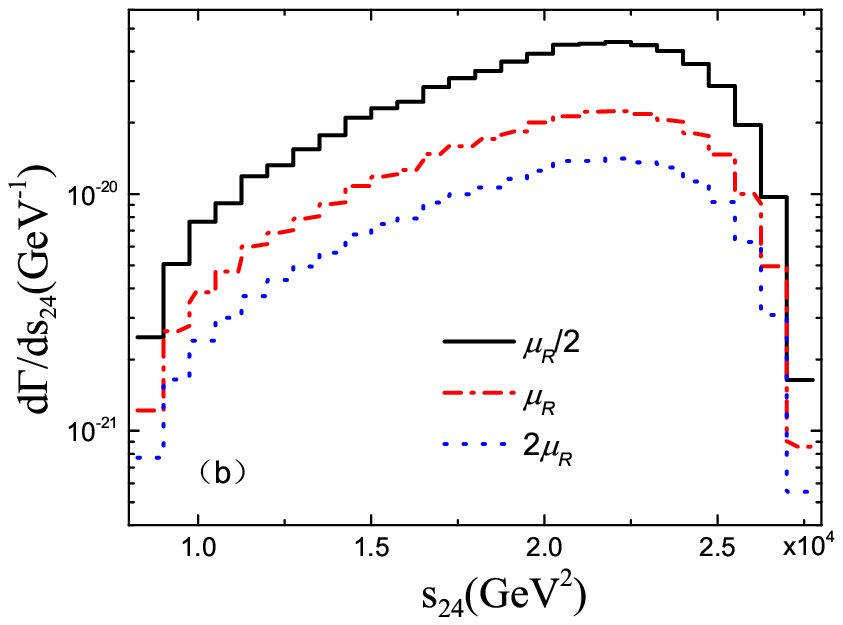}
\includegraphics[scale=0.6]{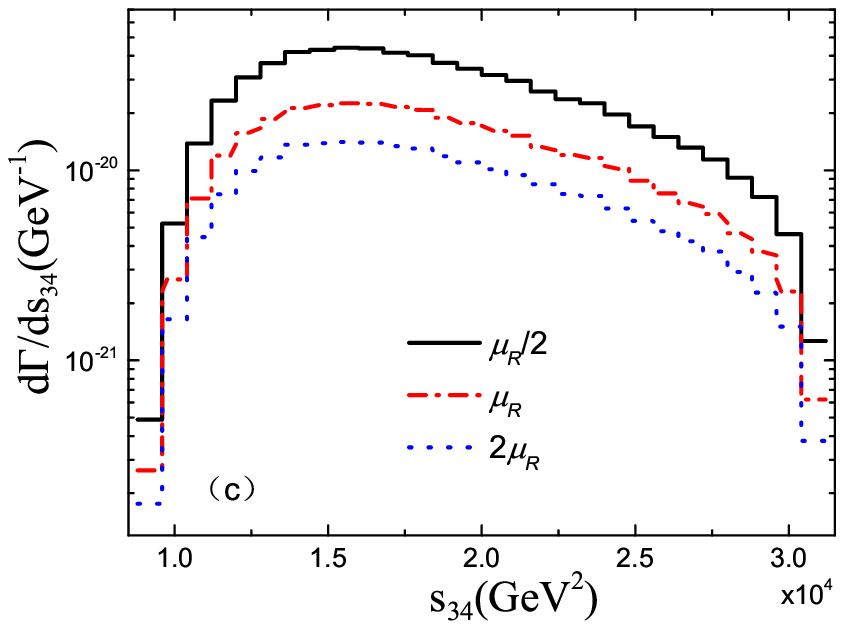}
\caption{The differential decay width $d\Gamma/ds_{23}$ (a), $d\Gamma/ds_{24}$ (b) and $d\Gamma/ds_{34}$ (c) with renormalization scale uncertainty for $ t\rightarrow |(c\bar{c})[n]\rangle +c+Z^{0}$.}
\label{alphacc}
\end{figure}

\begin{figure}[htb]
\includegraphics[scale=0.6]{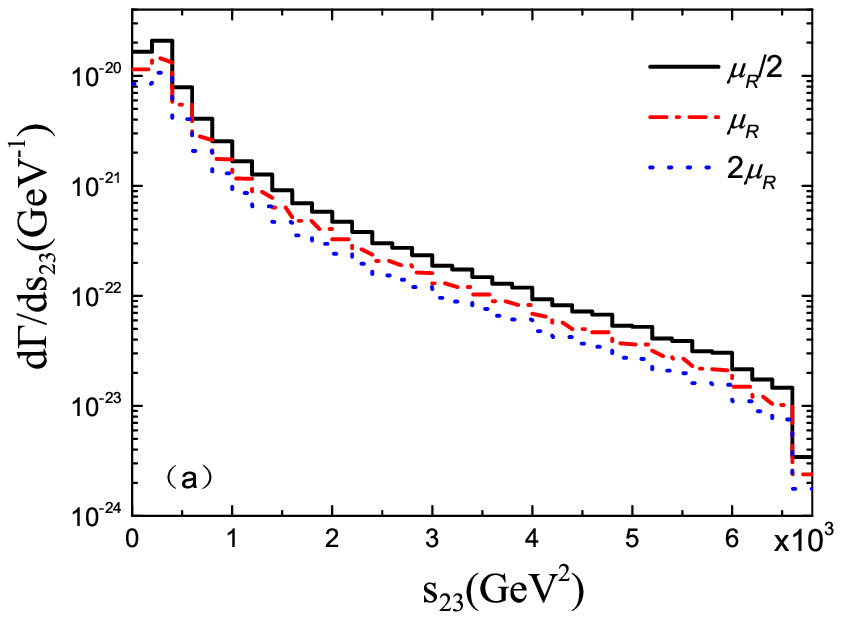}
\includegraphics[scale=0.6]{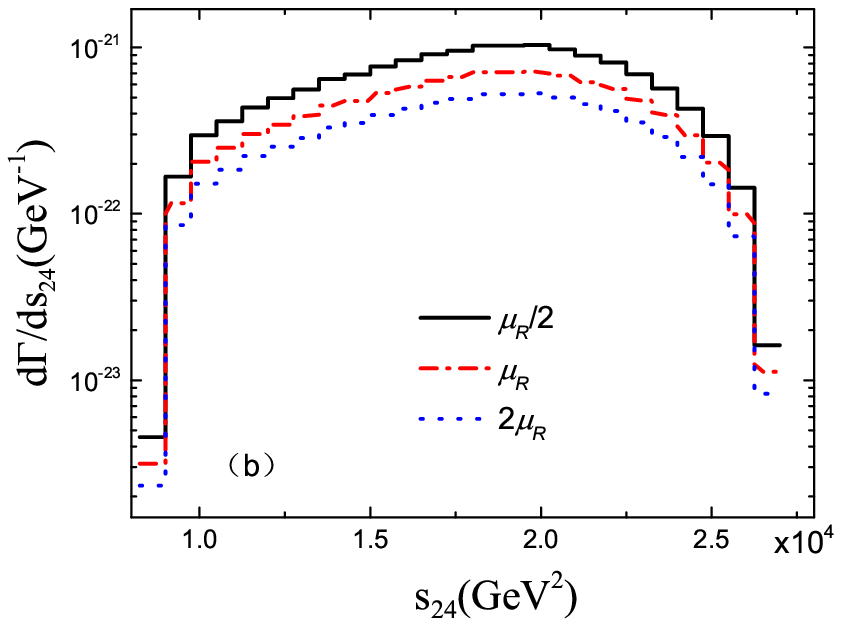}
\includegraphics[scale=0.6]{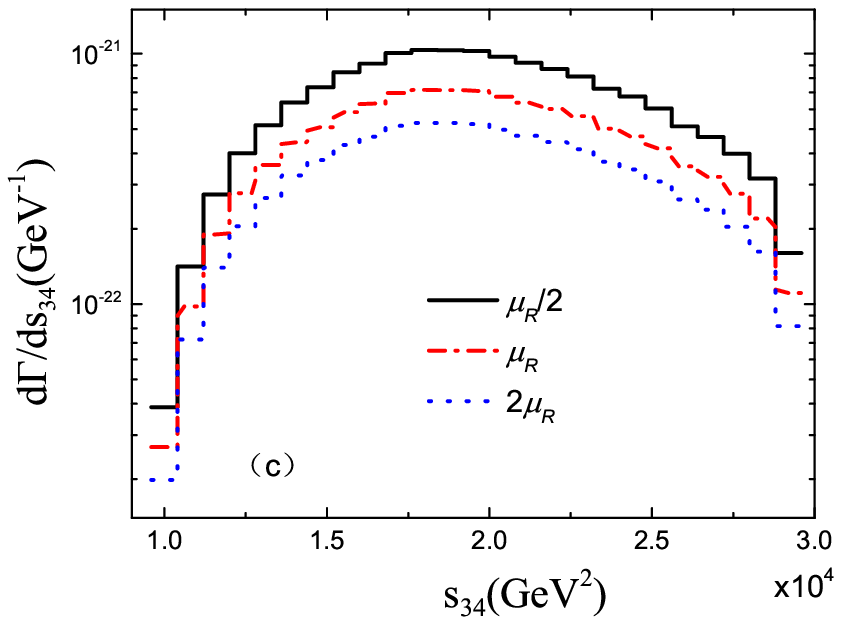}
\caption{The differential decay width $d\Gamma/ds_{23}$ (a), $d\Gamma/ds_{24}$ (b) and $d\Gamma/ds_{34}$ (c) with renormalization scale uncertainty for $ t\rightarrow |(c\bar{b})[n]\rangle +b +Z^{0}$.}
\label{alphacb}
\end{figure}

We present the scale uncertainties by varying the scale $\mu_R$ within the range of $[\mu_R/2,2\mu_R]$ in Table~\ref{tabmu}. Generally, the scale uncertainty can be suppressed by including high-order terms or by using an optimized scaling-setting method~\cite{Wu:2013ei, Wu:2014iba}. Here we set the renormalization scale to be $\mu_R=2m_c$ for charmonium production and $\mu_R=2m_b$ for $(c\bar{b})$-quarkonium production. Scale uncertainties for total invariant mass distributions are shown for the production of charmonium and $(c\bar{b})$-quarkonium in Figs.(\ref{alphacc}, \ref{alphacb}). Considering that the selected renormalization scale is small for the production of charmonium, the uncertainty is relatively larger than that for the production of $(c\bar{b})$-quarkonium.

\subsection{Background for the $(c\bar{b})$-quarkonium production}

\begin{figure}[htb]
\includegraphics[scale=0.51]{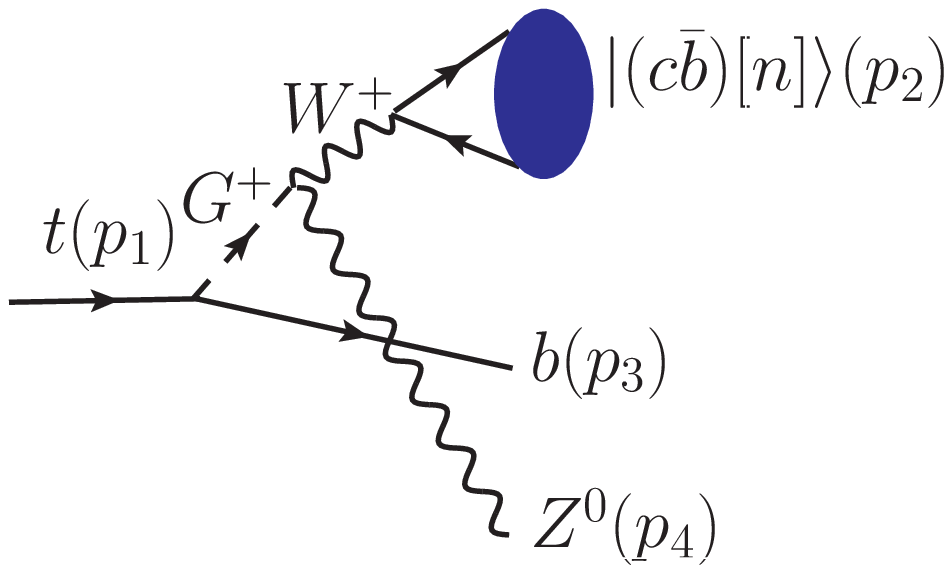}
\includegraphics[scale=0.51]{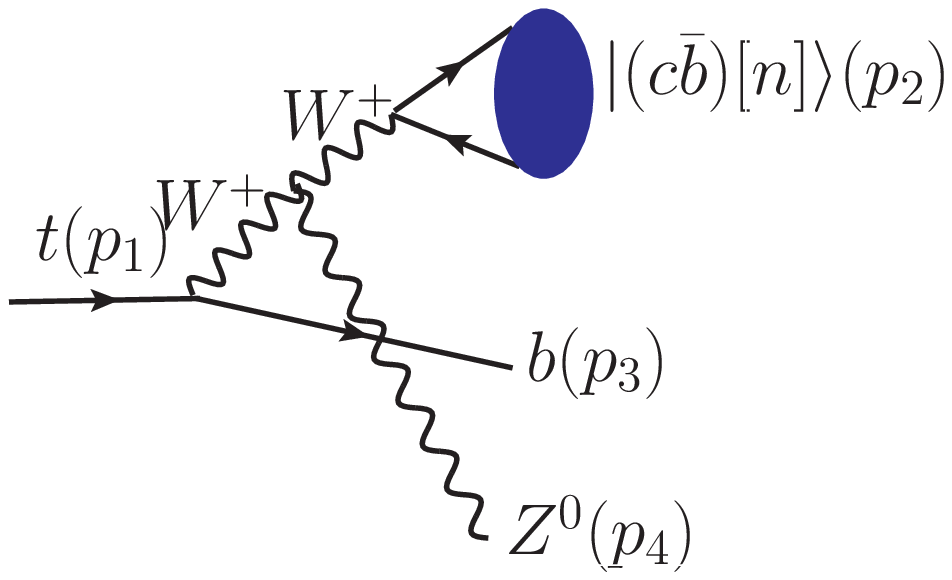}
\includegraphics[scale=0.51]{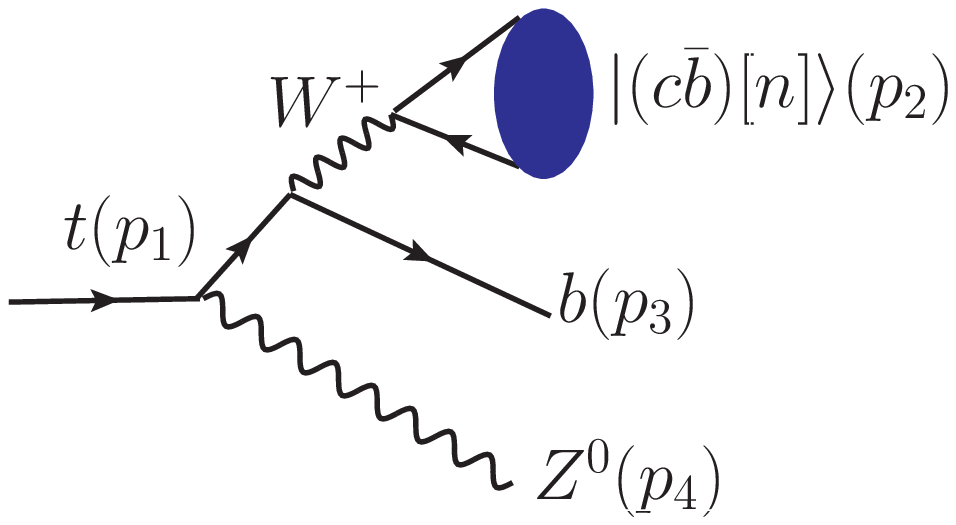}
\includegraphics[scale=0.51]{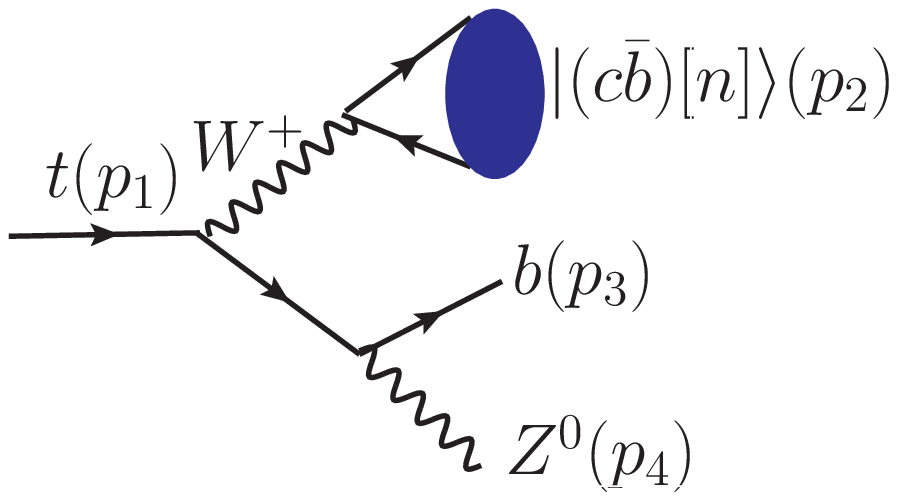}
\includegraphics[scale=0.51]{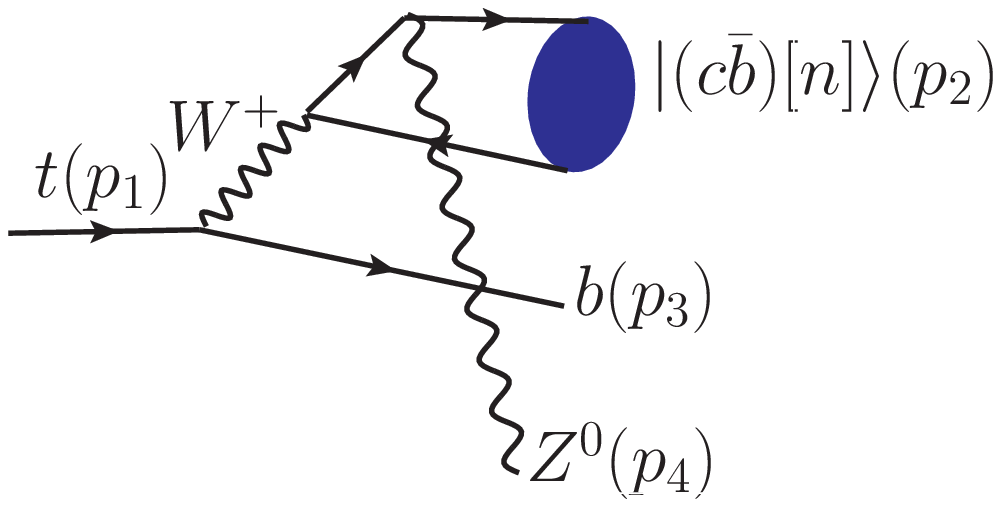}
\includegraphics[scale=0.51]{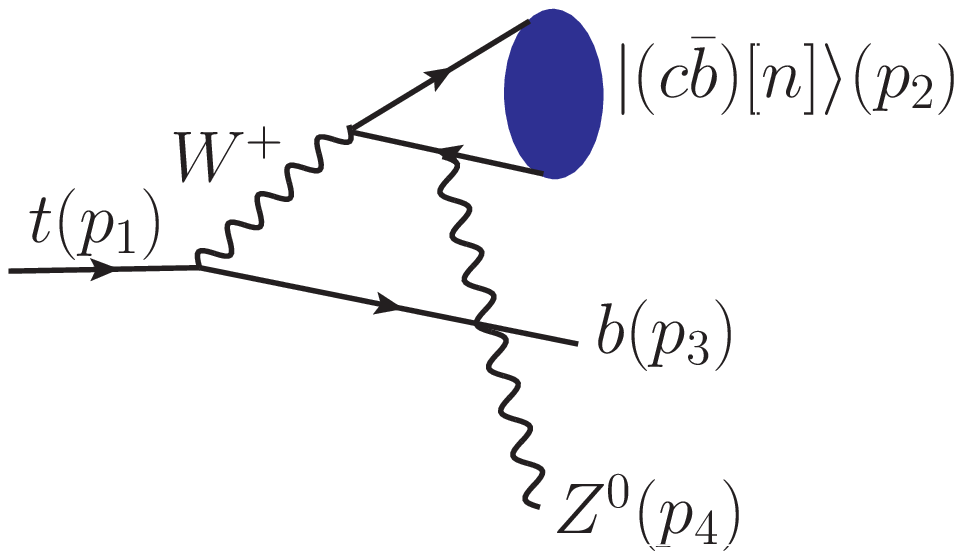}
\caption{The Feynman diagrams for $t(p_1)\to |(c\bar{b})[n]\rangle(p_2) + b(p_3) + Z^{0}(p_4)$ without FCNC, where $n$ stands for the two color-singlet $S$-wave states. }
\label{feyn2}
\end{figure}

For the production of $(c\bar{b})$-quarkonium with the same final states, there is another production channel, which could be treated as the background for observing the FCNC effect. The Feynman diagrams for the decay $t(p_1)\to |(c\bar{b})[n]\rangle(p_2) + b(p_3) + Z^{0}(p_4)$ without FCNC are plotted in Fig.(\ref{feyn2}), where $n$ stands for the two color-singlet $S$-wave states. For this channel, the short-distance amplitudes are
\begin{equation}
i M_{ss'}[n] = {\cal{C}}\; \sum\limits_{l = 11}^{16} {{\cal A}_l[n]},
\end{equation}
where $\cal{C}$ is $\frac{3\delta_{ij}}{\sqrt{3}}$ and the amplitudes ${\cal A}_{l}[n]$ are listed in Appendix B.

\begin{figure}[htb]
\includegraphics[scale=0.58]{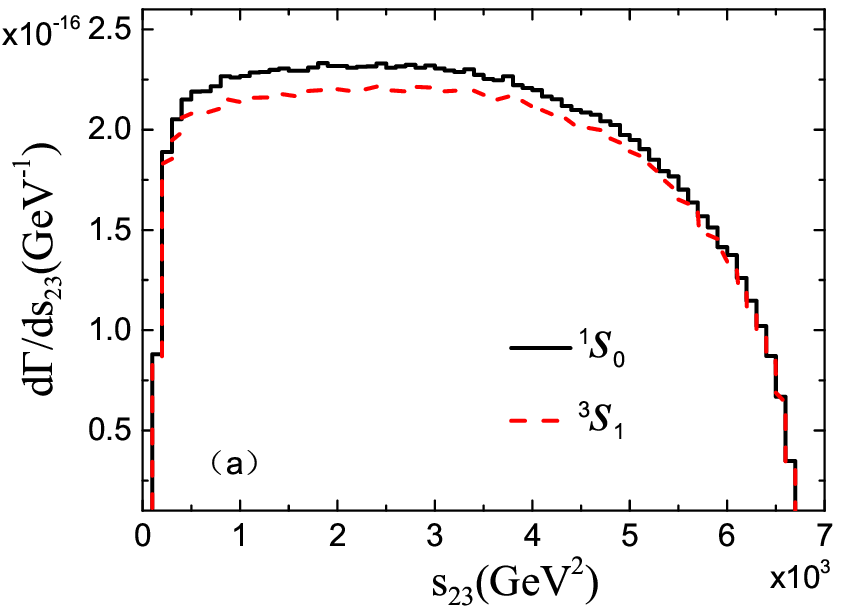}
\includegraphics[scale=0.6]{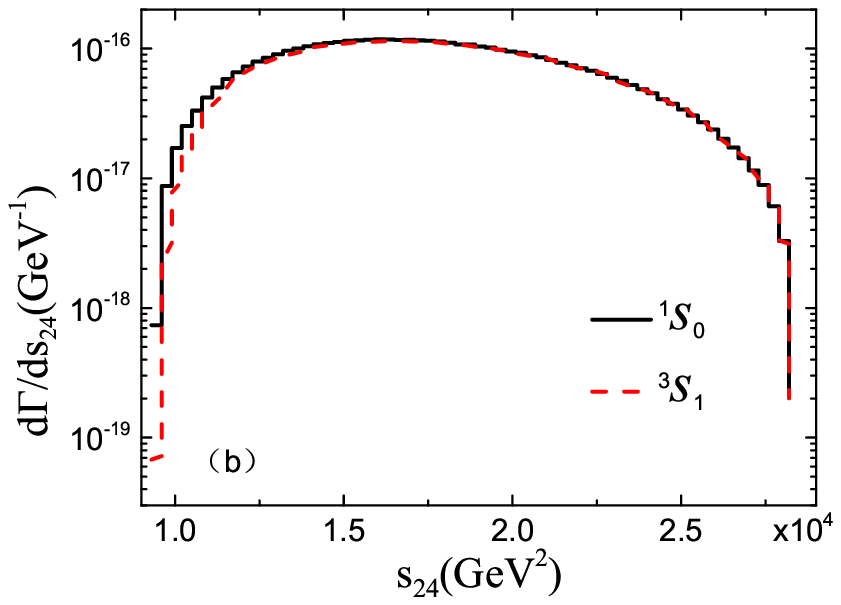}
\includegraphics[scale=0.6]{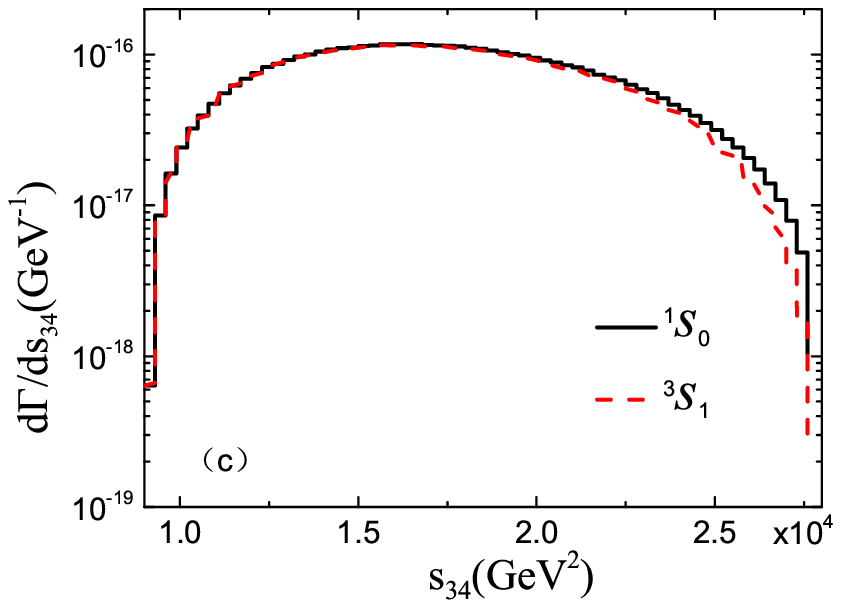}
\caption{The differential decay width not via FCNC $d\Gamma/ds_{23}$ (a), $d\Gamma/ds_{24}$ (b) and $d\Gamma/ds_{34}$ (c) for $ t\rightarrow |(c\bar{b})[n]\rangle +b+Z^{0}$.} \label{backtcbbzs}
\end{figure}

\begin{figure}[htb]
\includegraphics[scale=0.6]{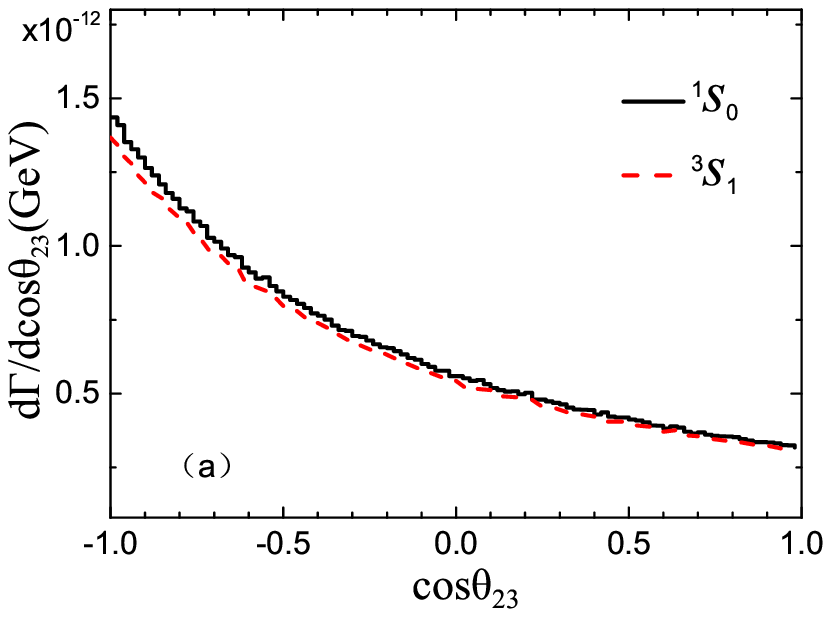}
\includegraphics[scale=0.6]{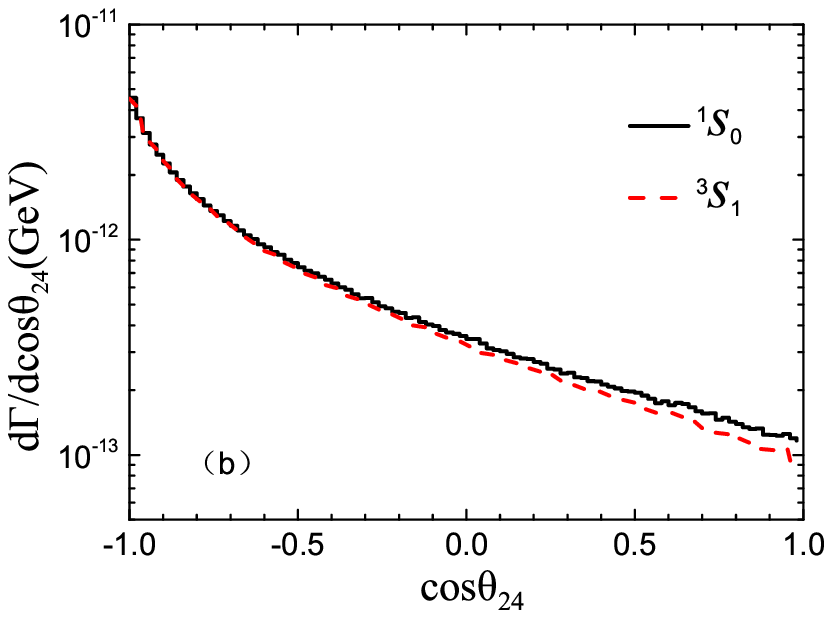}
\includegraphics[scale=0.6]{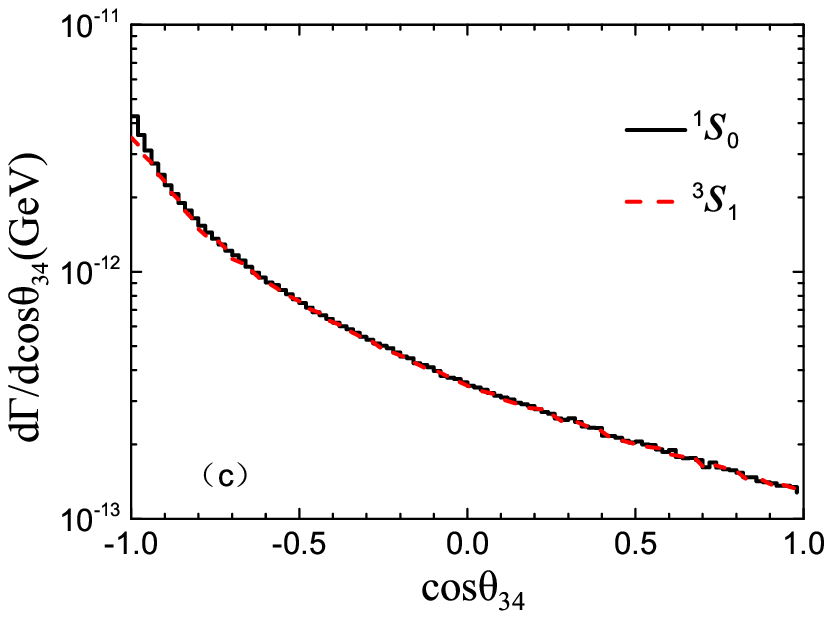}
\caption{The differential decay width not via FCNC $d\Gamma/d\cos\theta_{23}$ (a), $d\Gamma/d\cos\theta_{24}$ (b) and $d\Gamma/d\cos\theta_{34}$ (c) for $ t\rightarrow |(c\bar{b})[n]\rangle +b+Z^{0}$.} \label{backtcbbzcos}
\end{figure}

We present the invariant mass and the angular distributions for the production of $|(c\bar{b})[^1S_0]$ and $|(c\bar{b})[^3S_1]$ without FCNC in Figs.(\ref{backtcbbzs}, \ref{backtcbbzcos}). Figures.(\ref{dcos}, \ref{backtcbbzcos}) show the angular distributions $d\Gamma/d\cos\theta_{24}$ and $d\Gamma/d\cos\theta_{34}$ are close in shape, which the angular distribution $d\Gamma/d\cos\theta_{23}$ is quite different for the decay channels with or without FCNC. For example, the distribution $d\Gamma/d\cos\theta_{34}$ for the production without FCNC reaches its maximum value for $\theta_{23}=0$, while the distribution $d\Gamma/d\cos\theta_{34}$ for the production with FCNC reaches its maximum value for $\theta_{23}=1$. This difference is caused by the fact that for the production without FCNC, the quark components of $(c\bar{b})$-quarkonium are all from a off-shell $W^+$ boson. After integration, the total decay widths for the background process are $\Gamma(t\rightarrow B_c)=1.32\times 10^{-12}$ GeV and $\Gamma(t\rightarrow B^*_c)=1.26\times 10^{-12}$ GeV, respectively. They are larger than those of FCNC channels by about $10^5\sim10^6$ times, thus when searching of new physics signals from the FCNC channels, those background should be taken into consideration.

\subsection{New physics effects}

To simply estimate the new physics effects, we adopted $\Gamma=\Gamma_{t}\times BR(t\rightarrow cZ^0)\times R$, where $\Gamma_{t}$ is the total decay width of top quark about $2$ Gev, the related ratio R is given in subsection A and can be considered to be consistent with the SM on the order of magnitude. The branching ratio BR$(t\rightarrow cZ^0)$ has been studied in detail with many new models. Here we listed some estimated results in some new physics in Table~\ref{new}. We can find that the production of charmonium and $(c\bar{b})$-quarkonium through top quark decays may be accessible at LHC or HL-LHC running at $\sqrt{s}$ = 14~TeV and with the integrated luminosity of 3 $ab^{-1}$. 

\begin{table}[htb]
\begin{tabular}{|c||c|c|c|}
\hline new model    & ~BR$(t\rightarrow cZ^0)~$   & ~$\Gamma_{t\rightarrow (c\bar{c})+cZ^0}$~   & ~$\Gamma_{t\rightarrow (c\bar{b})+bZ^0}$~  \\
\hline \hline
2HDM type III & $10^{-3}$~\cite{Gaitan:2017tka}   & $10^{-7}$   & $10^{-9}$  \\
\hline
effective Lagrangian &$10^{-4}$~\cite{Shen:2018mlj}  & $10^{-8}$  &  $10^{-10}$  \\
\hline
models with extra quarks &$10^{-4}$~\cite{AguilarSaavedra:2002kr}  & $10^{-8}$  &  $10^{-10}$  \\
\hline
TC2 &$10^{-5}$~\cite{Lu:2003yr}  & $10^{-9}$  &  $10^{-11}$  \\
\hline
MSSM &$10^{-6}$~\cite{Larios:2006pb}  & $10^{-10}$  &  $10^{-12}$  \\
\hline
\end{tabular}
  \caption{The estimation of new physics effect with several new models.}
  \label{new}
\end{table}

\section{Summary}

The rare FCNC process is generally forbidden at the tree level in the SM, which is small and is used for searching of new physics beyond the SM. Within the framework of NRQCD, we have done a detailed study on the production of heavy-quarkonium through top quark semi-exclusive decays via FCNC, $t\to |(c\bar{Q})[n]\rangle +Q + Z^{0}$, where $Q$ stands for $c$ or $b$ quark, respectively. If assuming the spin-triplet $|(c\bar{Q})[^3S_1]\rangle$ decays to the ground $|(c\bar{Q})[^1S_0]\rangle$ with $100\%$ efficiency, the total decay width are as follows:
\begin{eqnarray}
\Gamma_{t\to |(c\bar{c})[^1S_0]\rangle} &=& 2.57^{+2.07+2.44}_{-1.02-0.96}\times 10^{-16}~{\rm GeV} , \\
\Gamma_{t\to |(c\bar{b})[^1S_0]\rangle} &=& 8.33^{+0.80+3.69}_{-0.79-2.18}\times 10^{-18}~{\rm GeV} ,
\end{eqnarray}
where the uncertainties from various quark masses and renormalization scales are summed up in quadrature. Various differential distributions have also been presented. Even though the decay widths are small, they are still important, which will provide useful guidance for searching of new physics beyond the SM from the heavy quarkonium involved processes. \\

{\bf Acknowledgements}: We would like to thank Xing-Gang Wu for useful discussion. This work was partially supported by the National Natural Science Foundation of China (No.11375008, No.11647307).

{\bf Appendix A}: The amplitudes $A_{l}$ of the process $t \to |(c\bar{Q})[n]\rangle + Q + Z^{0}$ via FCNC can be written as:
\begin{eqnarray}
\mathcal{A}_{1}=&&\int \frac{d ^4 q}{(2\pi)^4} (-i g_s)^2 \gamma_{\mu} \frac{\Pi_{p_2}[n]}{(p_3+p_{22})^2} \gamma_{\mu} \frac{\slashed{p}_{2}+\slashed{p}_3+m_c}{(p_2+p_3)^2-m_{c}^{2}}(ie)^3\frac{\gamma_{\nu}P_{L} \mathrm{CKM}(2,d_m)}{\sqrt{2}\sin{\theta_W}} \nonumber\\ &&\frac{\slashed{q}-\slashed{p}_4+m_{d_m}}{(q-p_{4})^2-m_{d_m}^{2}}
\left(\frac{\sin\theta_W\gamma_\eta P_{R}}{3\cos\theta_W}+\frac{\left(\frac{(\sin\theta_W)^2}{3}-\frac{1}{2}\right)\gamma_\eta P_{L}}{\cos\theta_W\sin\theta_W}\right)\slashed{\varepsilon}(p_4) \nonumber\\
&&\frac{\slashed{q}+m_{d_m}}{q^2-m_{d_m}^{2}} \frac{\gamma_{\nu}P_{L} \mathrm{CKM}(3,d_m)^\ast}{\sqrt{2}\sin{\theta_W}}\frac{1}{(q-p_2-p_3-p_4)^2-m_W^2} \nonumber\\
\mathcal{A}_{2}=&&-\frac{\cos{\theta_W}}{\sin{\theta_W}}\int \frac{d ^4 q}{(2\pi)^4} (-i g_s)^2 \gamma_{\mu} \frac{\Pi_{p_2}[n]}{(p_3+p_{22})^2} \gamma_{\mu} \frac{\slashed{p}_{2}+\slashed{p}_3+m_c}{(p_2+p_3)^2-m_{c}^{2}}(ie)^3\frac{\gamma_\alpha P_{L} \mathrm{CKM}(2,d_m)}{\sqrt{2}\sin{\theta_W}} \nonumber\\ &&\frac{\slashed{p}_2+\slashed{p}_3+\slashed{p}_4-\slashed{q}+m_{d_m}}{(q-p_2-p_3-p_4)^2-m_{d_m}^{2}} \frac{\gamma_\beta P_{L} \mathrm{CKM}(3,d_m)^\ast}{\sqrt{2}\sin\theta_W}\frac{\slashed{\varepsilon}(p_4)}{(q^2-m_W^2)((q-p_4)^2-m_W^2)}\nonumber\\
&&(g_{\alpha\beta}(p_4-2q)_\gamma+g_{\gamma\alpha}(q-2p_4)_\beta+g_{\gamma\beta}(p_4+q)_\alpha)\nonumber\\
\mathcal{A}_{3}=&&-\frac{m_W\sin\theta_W}{\cos\theta_W}\int \frac{d ^4 q}{(2\pi)^4} (-i g_s)^2 \gamma_{\mu} \frac{\Pi_{p_2}[n]}{(p_3+p_{22})^2} \gamma_{\mu} \frac{\slashed{p}_{2}+\slashed{p}_3+m_c}{(p_2+p_3)^2-m_{c}^{2}}(ie)^3\frac{ \gamma_{\nu}P_{L} \mathrm{CKM}(2,d_m)}{\sqrt{2}\sin{\theta_W}} \nonumber\\ &&\frac{\slashed{p}_2+\slashed{p}_3+\slashed{p}_4-\slashed{q}+m_{d_m}}{(q-p_2-p_3-p_4)^2-m_{d_m}^{2}}
\left(\frac{m_t P_{R} \mathrm{CKM}(3,d_m)^\ast}{\sqrt{2}m_W \sin\theta_W}-\frac{m_{d_m} P_{L} \mathrm{CKM}(3,d_m)^\ast }{\sqrt{2}m_W\sin\theta_W}\right) \nonumber\\
&&\frac{\slashed{\varepsilon}(p_4)}{(q^2-m_W^2)((q-p_4)^2-m_W^2)}\nonumber\\
\mathcal{A}_{4}=&&\int \frac{d ^4 q}{(2\pi)^4} (-i g_s)^2 \gamma_{\mu} \frac{\Pi_{p_2}[n]}{(p_3+p_{22})^2} (ie)^3\frac{ \gamma_{\nu}P_{L} \mathrm{CKM}(2,d_m)}{\sqrt{2}\sin{\theta_W}} \frac{-\slashed{q}+m_{d_m}}{q^2-m_{d_m}^{2}}\nonumber\\
&&\left(\frac{\sin\theta_W\gamma_\eta P_{R}}{3\cos\theta_W}+\frac{\left(\frac{(\sin\theta_W)^2}{3}-\frac{1}{2}\right)\gamma_\eta P_{L}}{\cos\theta_W\sin\theta_W}\right)\slashed{\varepsilon}(p_4) \nonumber\\
&&\frac{\slashed{p}_4-\slashed{q}+m_{d_m}}{(q-p_4)^2-m_{d_m}^{2}} \frac{\gamma_{\nu}P_{L} \mathrm{CKM}(3,d_m)^\ast}{\sqrt{2}\sin{\theta_W}} \frac{\slashed{p}_{21}+\slashed{p}_4+m_t}{(p_{21}+p_4)^2-m_{t}^{2}}\gamma_{\mu} \frac{1}{(q+p_{21})^2-m_W^2} \nonumber\\
\mathcal{A}_{5}=&&\frac{\cos{\theta_W}}{\sin{\theta_W}}\int \frac{d ^4 q}{(2\pi)^4} (-i g_s)^2 \gamma_{\mu} \frac{\Pi_{p_2}[n]}{(p_3+p_{22})^2} (ie)^3\frac{\gamma_\alpha P_{L} \mathrm{CKM}(2,d_m)}{\sqrt{2}\sin{\theta_W}} \nonumber\\
&&\frac{\slashed{p}_{21}+\slashed{q}+m_{d_m}}{(q+p_{21})^2-m_{d_m}^{2}} \frac{\gamma_\beta P_{L} \mathrm{CKM}(3,d_m)^\ast}{\sqrt{2}\sin\theta_W}\frac{\slashed{p}_{21}+\slashed{p}_4+m_t}{(p_{21}+p_4)^2-m_{t}^{2}}\nonumber\\
&& \gamma_{\mu} (g_{\alpha\beta}(p_4-2q)_\gamma+g_{\gamma\alpha}(q-2p_4)_\beta+g_{\gamma\beta}(q+p_4)_\alpha)\nonumber\\
&&\frac{\slashed{\varepsilon}(p_4)}{(q^2-m_W^2)((q-p_4)^2-m_W^2)} \nonumber\\
\mathcal{A}_{6}=&&-\frac{m_W\sin\theta_W}{\cos\theta_W}\int \frac{d ^4 q}{(2\pi)^4} (-i g_s)^2 \gamma_{\mu} \frac{\Pi_{p_2}[n]}{(p_3+p_{22})^2} (ie)^3\frac{\gamma_{\nu}P_{L} \mathrm{CKM}(2,d_m)}{\sqrt{2}\sin{\theta_W}} \nonumber\\
&&\frac{\slashed{p}_{21}+\slashed{q}+m_{d_m}}{(q+p_{21})^2-m_{d_m}^{2}} \left(\frac{m_t P_{R} \mathrm{CKM}(3,d_m)^\ast}{\sqrt{2}m_W \sin\theta_W}-\frac{m_{d_m} P_{L} \mathrm{CKM}(3,d_m)^\ast }{\sqrt{2}m_W\sin\theta_W}\right) \nonumber\\
&&\frac{\slashed{p}_{21}+\slashed{p}_4+m_t}{(p_{21}+p_4)^2-m_{t}^{2}}\gamma_{\mu} \frac{\slashed{\varepsilon}(p_4)}{(q^2-m_W^2)((q-p_4)^2-m_W^2)} \nonumber\\
\mathcal{A}_{7}=&&\int \frac{d ^4 q}{(2\pi)^4} (-i g_s)^2 \gamma_{\mu} \frac{\Pi_{p_2}[n]}{(p_3+p_{22})^2}(ie)^3\frac{\gamma_{\nu}P_{L} \mathrm{CKM}(2,d_m)}{\sqrt{2}\sin{\theta_W}} \frac{-\slashed{q}+\slashed{p}_{21}+m_{d_m}}{(q-p_{21})^2-m_{d_m}^{2}}\gamma_{\mu} \nonumber\\ &&\frac{-\slashed{q}+\slashed{p}_2+\slashed{p}_3+m_{d_m}}{(q-p_{2}-p_{3})^2-m_{d_m}^{2}}
\left(\frac{\sin\theta_W\gamma_\eta P_{R}}{3\cos\theta_W}+\frac{\left(\frac{(\sin\theta_W)^2}{3}-\frac{1}{2}\right)\gamma_\eta P_{L}}{\cos\theta_W\sin\theta_W}\right)\slashed{\varepsilon}(p_4) \nonumber\\
&&\frac{-\slashed{q}+\slashed{p}_2+\slashed{p}_3+\slashed{p}_4+m_{d_m}}{(q-p_{2}-p_{3}-p_{4})^2-m_{d_m}^{2}} \frac{\gamma_{\nu}P_{L} \mathrm{CKM}(3,d_m)^\ast}{\sqrt{2}\sin{\theta_W}}\frac{1}{q^2-m_W^2} \nonumber\\
\mathcal{A}_{8}=&&\int \frac{d ^4 q}{(2\pi)^4} (-i g_s)^2 \gamma_{\mu} \frac{\Pi_{p_2}[n]}{(p_3+p_{22})^2}(ie)^3\frac{\gamma_{\nu}P_{L} \mathrm{CKM}(2,d_m)}{\sqrt{2}\sin{\theta_W}} \frac{-\slashed{q}+\slashed{p}_{21}+m_{d_m}}{(q-p_{21})^2-m_{d_m}^{2}} \nonumber\\ &&\left(\frac{\sin\theta_W\gamma_\eta P_{R}}{3\cos\theta_W}+\frac{\left(\frac{(\sin\theta_W)^2}{3}-\frac{1}{2}\right)\gamma_\eta P_{L}}{\cos\theta_W\sin\theta_W}\right)\slashed{\varepsilon}(p_4) \frac{-\slashed{q}+\slashed{p}_{21}+\slashed{p}_4+m_{d_m}}{(q-p_{21}-p_{4})^2-m_{d_m}^{2}} \gamma_{\mu} \nonumber\\
&&\frac{-\slashed{q}+\slashed{p}_2+\slashed{p}_3+\slashed{p}_4+m_{d_m}}{(q-p_{2}-p_{3}-p_{4})^2-m_{d_m}^{2}} \frac{\gamma_{\nu}P_{L} \mathrm{CKM}(3,d_m)^\ast}{\sqrt{2}\sin{\theta_W}}\frac{1}{q^2-m_W^2} \nonumber\\
\mathcal{A}_{9}=&&-\frac{\cos{\theta_W}}{\sin{\theta_W}}\int \frac{d ^4 q}{(2\pi)^4} (-i g_s)^2 \gamma_{\mu} \frac{\Pi_{p_2}[n]}{(p_3+p_{22})^2} (ie)^3\frac{\gamma_\alpha P_{L} \mathrm{CKM}(2,d_m)}{\sqrt{2}\sin{\theta_W}} \frac{-\slashed{q}+\slashed{p}_{21}+\slashed{p}_4+m_{d_m}}{(q-p_{21}-p_4)^2-m_{{d_m}}^{2}} \nonumber\\ &&\gamma_{\mu}\frac{\slashed{p}_2+\slashed{p}_3+\slashed{p}_4-\slashed{q}+m_{d_m}}{(q-p_2-p_3-p_4)^2-m_{d_m}^{2}} \frac{\gamma_\beta P_{L} \mathrm{CKM}(3,d_m)^\ast}{\sqrt{2}\sin\theta_W}\frac{\slashed{\varepsilon}(p_4)}{(q^2-m_W^2)((q-p_4)^2-m_W^2)}\nonumber\\
&&(g_{\alpha\beta}(p_4-2q)_\gamma+g_{\gamma\alpha}(q-2p_4)_\beta+g_{\gamma\beta}(p_4+q)_\alpha)
\nonumber\\
\mathcal{A}_{10}=&&-\frac{m_W\sin\theta_W}{\cos\theta_W}\int \frac{d ^4 q}{(2\pi)^4} (-i g_s)^2 \gamma_{\mu} \frac{\Pi_{p_2}[n]}{(p_3+p_{22})^2} (ie)^3\frac{ \gamma_{\nu}P_{L} \mathrm{CKM}(2,d_m)}{\sqrt{2}\sin{\theta_W}} \frac{-\slashed{q}+\slashed{p}_{21}+\slashed{p}_4+m_{d_m}}{(q-p_{21}-p_4)^2-m_{d_m}^{2}} \nonumber\\ &&\gamma_{\mu} \frac{\slashed{p}_2+\slashed{p}_3+\slashed{p}_4-\slashed{q}+m_{d_m}}{(q-p_2-p_3-p_4)^2-m_{d_m}^{2}}
\left(\frac{m_t P_{R} \mathrm{CKM}(3,d_m)^\ast}{\sqrt{2}m_W \sin\theta_W}-\frac{m_{d_m} P_{L} \mathrm{CKM}(3,d_m)^\ast }{\sqrt{2}m_W\sin\theta_W}\right) \nonumber\\
&&\frac{\slashed{\varepsilon}(p_4)}{(q^2-m_W^2)((q-p_4)^2-m_W^2)}\nonumber
\end{eqnarray}
where $P_L=\frac{1-\gamma_5}{2}$, $P_R=\frac{1+\gamma_5}{2}$ and $d_m$ stands for the generation of down-type quark with mass $m_{d_m}$. The Cabibbo-Kobayashi-Maskawa (CKM) matrix $\mathrm{CKM}(2,3)=0.041$ and $\mathrm{CKM}(3,3)=1$.

{\bf Appendix B}: The amplitudes ${\cal A}_{l}[n]$ for the decay $t(p_1)\to |(c\bar{b})[n]\rangle(p_2) + b(p_3) + Z^{0}(p_4)$ without FCNC are:
\begin{eqnarray}
\mathcal{A}_{11}=&& -i\frac{m_W \sin\theta_W}{\cos\theta_W}(ie)^3
{\bar {u}_{s i}}({p_3})
\left(\frac{m_t P_{R}}{\sqrt{2}m_W \sin\theta_W}-\frac{m_{b} P_{L}}{\sqrt{2}m_W\sin\theta_W}\right)u_{s' j}({p_1})
\nonumber\\&&\frac{\slashed{\varepsilon}(p_4)}{(p_2+p_4)^2-m_{W}^{2}}
\mathrm{Tr}\left[\frac{\gamma_{\mu}P_{L} \mathrm{CKM}(2,3)}{\sqrt{2}\sin{\theta_W}}\frac{\Pi_{p_2}[n]}{p_2^2-m_W^2}\right]\nonumber\\
\mathcal{A}_{12}=&& i\frac{\cos\theta_W}{\sin\theta_W}(ie)^3
{\bar {u}_{s i}}({p_3})
\frac{\gamma_{\mu}P_{L}}{\sqrt{2}\sin{\theta_W}}u_{s' j}({p_1})
\frac{\slashed{\varepsilon}(p_4)}{(p_2+p_4)^2-m_{W}^{2}}
\mathrm{Tr}\left[\frac{\gamma_{\nu}P_{L} \mathrm{CKM}(2,3)}{\sqrt{2}\sin{\theta_W}}\frac{\Pi_{p_2}[n]}{p_2^2-m_W^2}\right]
\nonumber\\&&(g_{\alpha\mu}(-2p_4-p_2)_\nu+g_{\alpha\nu}(p_4-p_2)_\mu+g_{\mu\nu}(2p_2+p_4)_\alpha)\nonumber\\
\mathcal{A}_{13}=&&i(ie)^3{\bar {u}_{s i}}({p_3})\frac{\gamma_{\nu}P_{L}}{\sqrt{2}\sin{\theta_W}}
\frac{\slashed{p}_{2}+\slashed{p}_3+m_t}{(p_2+p_3)^2-m_{t}^{2}}\left(\frac{\left(\frac{1}{2}-\frac{2(\sin\theta_W)^2}{3}\right)\gamma_\mu P_{L}}{\cos\theta_W\sin\theta_W}-\frac{2\sin\theta_W\gamma_\mu P_{R}}{3\cos\theta_W}\right)\nonumber\\&&\slashed{\varepsilon}(p_4)u_{s' j}({p_1})\mathrm{Tr}\left[\frac{\gamma_{\nu}P_{L} \mathrm{CKM}(2,3)}{\sqrt{2}\sin{\theta_W}}\frac{\Pi_{p_2}[n]}{p_2^2-m_W^2}\right]\nonumber\\
\mathcal{A}_{14}=&& i(ie)^3 {\bar {u}_{s i}}({p_3})\left(\frac{\sin\theta_W\gamma_\eta P_{R}}{3\cos\theta_W}+\frac{\left(\frac{(\sin\theta_W)^2}{3}-\frac{1}{2}\right)\gamma_\eta P_{L}}{\cos\theta_W\sin\theta_W}\right)\slashed{\varepsilon}(p_4)\frac{\slashed{p}_{3}+\slashed{p}_4+m_b}{(p_3+p_4)^2-m_{b}^{2}} \nonumber\\&&\frac{\gamma_{\mu}P_{L}}{\sqrt{2}\sin{\theta_W}}{u_{s' j}}({p_1})\mathrm{Tr}\left[\frac{\gamma_{\mu}P_{L} \mathrm{CKM}(2,3)}{\sqrt{2}\sin{\theta_W}}\frac{\Pi_{p_2}[n]}{p_2^2-m_W^2}\right]\nonumber\\
\mathcal{A}_{15}=&&i(ie)^3 {\bar {u}_{s i}}({p_3})\frac{\gamma_{\mu}P_{L}}{\sqrt{2}\sin{\theta_W}}{u_{s' j}}({p_1})\mathrm{Tr}\left[\left(\frac{\left(\frac{1}{2}-\frac{2(\sin\theta_W)^2}{3}\right)\gamma_\nu P_{L}}{\cos\theta_W\sin\theta_W}-\frac{2\sin\theta_W\gamma_\nu P_{R}}{3\cos\theta_W}\right)\slashed{\varepsilon}(p_4)\right.\nonumber\\
&&\left.\frac{\slashed{p}_{21}+\slashed{p}_4+m_c}{(p_{21}+p_4)^2-m_{c}^{2}}
\frac{\gamma_{\mu}P_{L}\mathrm{CKM}(2,3)}{\sqrt{2}\sin{\theta_W}}
\frac{\Pi_{p_2}[n]}{(p_2+p_4)^2-m_W^2}\right]\nonumber\\
\mathcal{A}_{16}=&&i(ie)^3 {\bar {u}_{s i}}({p_3})\frac{\gamma_{\mu}P_{L}}{\sqrt{2}\sin{\theta_W}}{u_{s' j}}({p_1})\mathrm{Tr}\left[\frac{\gamma_{\mu}P_{L} \mathrm{CKM}(2,3)}{\sqrt{2}\sin{\theta_W}}\frac{\Pi_{p_2}[n]}{(p_2+p_4)^2-m_W^2}\right.\nonumber\\
&&\left.\frac{-\slashed{p}_{22}-\slashed{p}_4+m_b}{(p_{22}+p_4)^2-m_{b}^{2}}\left(\frac{\left(\frac{(\sin\theta_W)^2}{3}-\frac{1}{2}\right)\gamma_\nu P_{L}}{\cos\theta_W\sin\theta_W}+\frac{\sin\theta_W\gamma_\nu P_{R}}{3\cos\theta_W}\right)\slashed{\varepsilon}(p_4)\right]\nonumber
\end{eqnarray}

\end{document}